\documentclass[journal,comsoc,twocolumn]{IEEEtran}
\IEEEoverridecommandlockouts
\usepackage{amsmath,amsfonts}
\usepackage{amssymb}
\usepackage[ruled]{algorithm2e}
\usepackage[noend]{algpseudocode}
\usepackage{array}
\usepackage{algorithmicx}
\usepackage{threeparttable}
\usepackage{multirow}
\usepackage{diagbox}
\usepackage{subfigure}
\usepackage{multicol}                 
\usepackage{multirow}                
\usepackage{float}                    
\usepackage{makecell}                 
\usepackage{booktabs}
\usepackage{utfsym}
\usepackage{textcomp}
\usepackage{stfloats}
\usepackage{url}
\usepackage{verbatim}
\usepackage{graphicx}
\usepackage{cite}

\begin{document}

\title{\makebox[\linewidth]{\parbox{\dimexpr\textwidth+2cm\relax}{\centering Auto-CsiNet: Scenario-customized Automatic \\Neural Network Architecture Generation for Massive MIMO CSI Feedback}}}
	
\author{
\normalsize {Xiangyi~Li,
Jiajia~Guo,
\IEEEmembership{\normalsize {Member,~IEEE}},\\
Chao-Kai~Wen, \IEEEmembership{\normalsize {Fellow,~IEEE}},
and
Shi~Jin, \IEEEmembership{\normalsize {Fellow,~IEEE}}
}
\thanks{X.~Li, J.~Guo, and S.~Jin are with the National Mobile Communications Research Laboratory, Southeast University, Nanjing, 210096, P. R. China (email: Xiangyi\_li@seu.edu.cn, jiajiaguo@seu.edu.cn jinshi@seu.edu.cn).}
\thanks{C.-K.~Wen is with the Institute of Communications Engineering, National Sun Yat-sen University, Kaohsiung 80424, Taiwan (e-mail: chaokai.wen@mail.nsysu.edu.tw).}
}

\maketitle

\begin{abstract}

Deep learning has revolutionized the design of the channel state information (CSI) feedback module in wireless communications. However, designing the optimal neural network (NN) architecture for CSI feedback can be a laborious and time-consuming process. Manual design can be prohibitively expensive for customizing NNs to different scenarios. This paper proposes using neural architecture search (NAS) to automate the generation of scenario-customized CSI feedback NN architectures, thereby maximizing the potential of deep learning in exclusive environments. By employing automated machine learning and gradient-descent-based NAS, an efficient and cost-effective architecture design process is achieved. The proposed approach leverages implicit scene knowledge, integrating it into the scenario customization process in a data-driven manner, and fully exploits the potential of deep learning for each specific scenario. To address the issue of excessive search, early stopping and elastic selection mechanisms are employed, enhancing the efficiency of the proposed scheme. The experimental results demonstrate that the automatically generated architecture, known as Auto-CsiNet, outperforms manually-designed models in both reconstruction performance (achieving approximately a 14\% improvement) and complexity (reducing it by approximately 50\%). Furthermore, the paper analyzes the impact of the scenario on the NN architecture and its capacity.

\end{abstract}



\section{Introduction}

Massive multiple-input and multiple-output (MIMO) is crucial for 6th generation (6G) wireless communication networks, as it enables high throughput, multiple streams, and pervasive coverage for diverse applications in smart cities \cite{6GMIMO,Lulu2014Liye}.
However, the acquisition of downlink channel state information (CSI) for high-quality precoding in massive MIMO systems is challenging due to the low channel reciprocity of frequency division duplexing (FDD) systems \cite{marzetta2015massive,Chataut2010MIMO}. To reduce feedback overhead, compressing CSI at the user equipment (UE) and reconstructing it at the base station (BS) is necessary \cite{love2008overview}. Traditional CSI compression and feedback methods, such as compressed sensing (CS) \cite{liang2020deep} and codebook-based methods \cite{Ra2015codebook,Dreifuerst2023Machine}, have limitations due to the complexity of the iterative algorithm and sophisticated codebook design. Therefore, new enabling technology is urgently needed to handle the high-dimensional nonlinear problem of CSI feedback.

The rapid development of artificial intelligence (AI) has been fueled by substantial improvements in hardware computing power, mass data collection, and device storage capacity, creating infinite possibilities for intelligent communications \cite{liu2021toward,Toward6G2021,dorner2017deep,Guiguan2020}. Motivated by the excellent performance of deep learning (DL), intelligent CSI feedback methods \cite{wen2018deep} are far superior to traditional algorithms in terms of performance and speed \cite{guo2022overview}, thus attracting widespread attention from the academic community. The key idea behind the DL-based CSI feedback scheme is to use neural network (NN) to achieve effective compression and decompression of CSI, which includes an encoder network at the UE for dimensional reduction of the high-dimensional CSI matrix. The encoder network outputs the compressed code before feedbacking to the BS, and a decoder network reconstructs the original CSI from the compressed code. The data-driven DL can quickly fit solutions via deep NN and achieve efficient and highly accurate reconstruction through the combination of ``offline training" and ``online deployment" modes.

One key issue in DL-based designs for wireless communication is constructing an efficient NN to maximize performance in a specific scenario. Novel NN architectures can be categorized into architectures based on convolutional neural networks (CNN), attention mechanisms \cite{cui2022transnet,cai2019attention,zhang2022attention}, generative models \cite{tolba2020massive,hussien2022prvnet}, and feature preprocessing and extraction \cite{ji2021clnet,sun2021lightweight,ma2021model}. This article focuses mainly on CNN-based architectures, which have advanced in four ways: (1) receptive field amplification\cite{guo2020convolutional, wang2021multi,lu2020multi}, (2) multiple resolutions\cite{lu2020multi,tang2022dilated,hu2021mrfnet,xu2021dfecsinet}, (3) lightweight convolutions\cite{tang2022dilated,cao2021lightweight}, and (4) flexible information interaction\cite{yu2020ds,song2021saldr}. The decoder network typically employs a cell/block-based structure, e.g., the RefineNet block, with a multi-branch (multiple resolutions) structure within the cell. Notably, most CNN-based architectures for CSI feedback share these two characteristics. Table \ref{Tab:NN-archi-performance} summarizes the number of cells and branches, and whether the cell contains skip connections.

However, these architectures are all manually designed, which involves a substantial and laborious workload in adjusting architecture hyperparameters. Moreover, most of these designs have focused on improving NN performance on the COST2100 indoor/outdoor scene dataset (released by \cite{wen2018deep}). There is no guarantee that these manually-designed network structures will perform equally well when applied to other scene CSI datasets.
Designing NNs manually faces two main challenges:
\begin{enumerate}
    \item Artificial NN design is an inefficient ``trial and error'' process, leading to high costs, such as time, manpower, computing power, and other resource consumption. The workload of manual design in adjusting the hyperparameters of NN architectures is also tremendous, laborious, and repetitive. Moreover, expert experience and knowledge are crucial in manual design, which requires manual attention in each link and cannot be realized in automatic mode. Constrained by experience and thinking, the design structure easily falls into the local optimal solution.

    \item The CSI feedback module requires different NN architectures for various scenarios and tasks, influenced by factors such as scatterer density, mobile terminal speed, antenna distribution, and bandwidth. A general NN may not achieve optimal performance for a specific task, as observed in Table \ref{Tab:NN-archi-performance} where the best architectures for indoor and outdoor scenarios are not the same. Manually designing task-oriented NN structures is costly, and the development of custom-designed NNs for practical applications has limitations. Thus, DL cannot fully realize its potential in specific scenarios.
\end{enumerate}


This work concentrates on the design of a scenario-customized NN architecture for intelligent CSI feedback. To overcome the challenges of labor-intensive manual design and the prohibitive costs associated with scenario customization, we introduce Auto-CsiNet. This automatic design scheme employs NN architecture search (NAS) and autonomous machine learning (AutoML) \cite{he2021automl,karmaker2021automl} to generate a structure for the CSI feedback NN without the need for extensive labor resources or expertise.

Furthermore, this economical and automatic design scheme not only accomplishes scenario-customized design of NN architectures but also maximizes the performance of Auto-CsiNet for specific tasks, thereby unlocking the full potential of deep learning in the CSI feedback module. Through AutoML, scene knowledge is seamlessly and implicitly integrated into the customization process in a data-driven manner. This enables manufacturers to efficiently acquire customized networks by adopting NAS, achieving complete automation and standardization in industrial processes.

Auto-CsiNet is based on CS-CsiNet \cite{wen2018deep} and is designed by modifying the RefineNet cell searched using NAS. The PC-DARTS method \cite{PC-DARTS}, a state-of-the-art and efficient NAS method, is utilized, enabling Auto-CsiNet to maintain a low design cost in terms of computing power and time cost, and making it easy to implement in the industry. Simulation results demonstrate that Auto-CsiNet outperforms numerous manually designed networks in reconstruction accuracy and NN complexity.
The major contributions of this work can be summarized as follows:
\begin{itemize}
    \item \textbf{The necessity of scenario-specific customization:}
    The investigation focuses on the problem of limited generalization of NN architectures and analyzes the cause, including the relationship between the capacity of NNs and the entropy of CSI information, the balance between the performance of NNs and transplantation, and the difficulty in network convergence.

    \item \textbf{The automatic generation framework:}
    Auto-CsiNet, a novel framework for the automatic generation of the CSI feedback NN architecture, is proposed. Auto-CsiNet utilizes AutoML and the low-cost NAS method to reduce designing cost and enable a labor-free and standardized designing process that is friendly to manufacturers. Additionally, Auto-CsiNet can tap into the maximum potential of NN architectures to achieve optimal performance on specific scenarios.

    \item \textbf{Refinements in evaluation criteria:}
    As the appropriate number of search rounds vary with different scenarios or feedback NN settings and over-searching can lead to degraded performance of automatically generated CSI feedback NNs, the evaluation criteria are refined by introducing early stop and elastic selection mechanisms. These refinements further ensure high performance in CSI feedback applications.

\end{itemize}

\par The rest of this paper is organized as follows: Section \ref{section:system model} presents the system model and the relationship between the NN architectures and scenario characteristics. Section \ref{section:framework} displays the proposed automatic generation framework of the CSI feedback NN structures based on AutoML, including the mechanism of NAS and the complete framework with the advanced evaluating criterion. Section \ref{section:Simulation-Results} provides CSI simulation details and the evaluation of our proposed Auto-CsiNet compared with the manually designed NNs. Section \ref{section:conclusion} gives the concluding remarks.


\par \emph{Notations:} Vectors and matrices are denoted by boldface lower and upper case letters, respectively. $(\cdot)^{\mathrm{H}}$ and $\sf{cov}(\cdot)$ denote Hermitian transpose and the covariance, respectively. $\mathbb{C}^{m\times n}$ or $\mathbb{R}^{m\times n}$ denotes the space of $m\times n$ complex-valued or real-valued matrix. For a 2-D matrix $\mathbf{A}$, $\mathbf{A}[i,j]$, $\sf{row}_k(\mathbf{A})$ and $\sf{col}_k(\mathbf{A})$ represent the $(i,j)^{\rm th}$ element, the $k^{\rm th}$ row and $k^{\rm th}$ column in matrix $\mathbf{A}$. $\mathsf{vec}(\mathbf{A})$ means the vectorized $\mathbf{A}$. For a 1-D vector $\mathbf{a}$, $\mathbf{a}[i]$ denotes its $i^{\rm th}$ element. $|\cdot|$ measures the scale of a set and $\|\cdot\|_2$ is the Euclidean/L2 norm. $[a_k]_{k=1}^N$ represents a list of $[a_1,a_2,...,a_N]$. $\sf{para}(\cdot)$ and $\sf{FLOPs}(\cdot)$ denote the NN's parameter amount and the floating-point operations (FLOPs) amount. $\mathrm{C}_n^m={n!}/{(m!(n-m)!)}$ stands for composite number.

\begin{table*}[t]
    \renewcommand\arraystretch{0.8}
    \caption{\label{Tab:NN-archi-performance}
    Performance, complexity, and architecture characteristic of manual-designed CNN-based autoencoder networks.}
    \centering
    \setlength{\abovecaptionskip}{0.0cm}
    \setlength{\belowcaptionskip}{0.0cm}
    \resizebox{0.85\textwidth}{!}{

    \begin{tabular}{c|ccccccc}\toprule
                          \multirow{2}{*}{CR=1/4}   & \multicolumn{2}{c}{NMSE [dB]}       & \multicolumn{2}{c}{Complexity}

                         & Cell number
                         & Branch
                         & Skip \\
                         & Indoor     & Outdoor & FLOPs [M] & Para [M] & in decoder & number & Connection\\ \midrule
    CsiNet\cite{wen2018deep}     & -17.36 & -8.75 & 5.42 & 2.10  &2 & 1 &$\usym{1F5F8}$\\
Attention-CsiNet\cite{cai2019attention} & -20.29 & -10.43 & 24.72 & $\backslash$   &2 & 1 &$\usym{1F5F8}$\\
DS-NLCsiNet\cite{yu2020ds} & -24.99 & -12.09 & 5.75 & 2.11  &2 & 1 &Densely connected\\
CsiNet+\cite{guo2020convolutional} & -27.37& -12.40 & 24.79 & 2.12  &5 & 1 &$\usym{1F5F8}$\\
CRNet\cite{lu2020multi} & -26.99 & -12.72 & 5.12& 2.10  &2 & 2 &$\usym{1F5F8}$\\
MRFNet\cite{hu2021mrfnet}    & -25.76& \textbf{-15.95} & 660 & 4.04 &3 & 3&$\usym{1F5F8}$\\
DCRNet\cite{tang2022dilated} &\textbf{-30.61} &-13.72 &17.57 &2.12  &2 & 2&$\usym{1F5F8}$\\
DFECsiNet\cite{xu2021dfecsinet}&-27.50 &-12.25 &6.38 &2.10  &2 & 2&$\usym{1F5F8}$\\

                         \bottomrule
    \end{tabular}}
\end{table*}

\section{System model and Problem formulation }\label{section:system model}
\subsection{Massive MIMO-OFDM FDD System}
\label{MIMO system model1}
Consider a typical urban micro-cell in FDD massive MIMO system with one BS serving for multiple single-antenna UEs. The BS is placed in the center of the cell and equipped with an $N_{\mathrm t}$-antenna uniform linear array (ULA)\footnote{We adopt the ULA model here for simpler illustration, while the analysis and the proposed model are not restricted to any specific array shape.}. We apply orthogonal frequency division multiplexing (OFDM) in downlink transmission over $N_{\rm f}$ subcarriers. The received signal on the $n^{\rm th}$ subcarrier for a UE can then be modeled as \cite{wen2018deep}:
\begin{equation}
y_n = {\mathbf h}_n^{\mathrm{H}}{\mathbf v}_{n}x_n  + z_{n},
\label{eq1}
\end{equation}
where ${\mathbf h}_n \in \mathbb{C}^{N_{\mathrm t}}$,  $x_{n} \in \mathbb{C}$ and  $z_{n} \in \mathbb{C}$ denote the downlink instantaneous channel vector in the frequency domain, the transmit data symbol and the additive noise, respectively. The beamforming or precoding vector $ {\mathbf v}_n \in \mathbb{C}^{N_{\mathrm t}}$
should be designed by the BS based on the received downlink CSI. In this paper,
we assume that perfect CSI has been acquired through pilot-based channel estimation and focus on the design of feedback approaches.
We stack all the $N_{\mathrm f}$ frequency channel vectors and derive the downlink CSI metrix ${\mathbf H}_{\mathrm {SF}}=[{\mathbf h}_1,{\mathbf h}_2,...,{\mathbf h}_{N_{\mathrm f}}]\in \mathbb{C}^{N_{\mathrm t}\times N_{\mathrm f}}$ in the spatial-frequency domain.
\par We represent the channel matrices in the angular-delay
domain using a 2D discrete Fourier transform (DFT) to better display the feature sparsity, which can be described as following:
\begin{equation}
\widetilde{{\mathbf H}}_{\mathrm {AD}}={\mathbf F}_{\mathrm s}{\mathbf H}_{\mathrm {SF}}{\mathbf F}_{\mathrm f},
\label{eq2}
\end{equation}
where ${\mathbf F}_{\mathrm s}$ and ${\mathbf F}_{\mathrm f}$ are $N_{\mathrm t}\times N_{\mathrm t}$ and $N_{\mathrm f}\times N_{\mathrm f}$ DFT matrices, respectively.
The high-dimensional CSI matrix should be sparse in the delay domain, and only the front $N_{\mathrm c}$ ($N_{\mathrm c}<N_{\mathrm f}$) rows displays distinct non-zero values, because the time delay among multiple paths only exists in a particularly limited period of time.
Hence, we retain the first $N_{\mathrm c}$ non-zero rows to further reduce the feedback overload and derive the dimension-reduced CSI in the angle-delay domain: ${\mathbf H}_{\mathrm {AD}} \in \mathbb{C}^{N_{\mathrm t}\times N_{\mathrm c}}$.
Moreover, the channel matrix is also sparse in a defined angle domain if the number of the transmit $N_{\mathrm t}$ tends to infinite.

Notice that in this paper, all statistical analysis of the angle-delay CSI are for ${\mathbf H}_{\mathrm {AD}}$. The complex-valued elements in ${\mathbf H}_{\mathrm {AD}}$ are also divided into real and imaginary real-valued parts, then normalized in the range of [0,1], where we finally obtain the NN's input: ${\mathbf H}\in \mathbb{R}^{N_{\mathrm t}\times N_{\mathrm c}\times 2}$.

\subsection{Single Side DL-based CSI Feedback}
The whole DL-based CSI feedback process is realized by
a framework combining CS and DL, the same as CS-CsiNet \cite{wen2018deep}, which consists a random linear projection at UE for CSI matrix dimension reduction and a decoder NN at BS for CSI reconstruction. The CS-based compression process can be expressed as:
\begin{equation}
    \mathbf{s}=\mathbf{A}\mathsf{vec}(\mathbf{H}),
    \label{equ3}
\end{equation}
where $\mathsf{vec}(\mathbf{H})\in \mathbb{R}^{N}$ ($N=2N_{\mathrm t}\cdot N_{\mathrm c}$) is composed of the real and imaginary parts of the vectored ${\mathbf H}\in \mathbb{R}^{N_{\mathrm t}\times N_{\mathrm c}\times 2}$, and $\mathbf{s}\in \mathbb{R}^{M}$ ($M<N$) presents the compressed codeword, where $M$ is determined by the predefined compression ratio (CR) that $M=N\times \mathrm{CR}$. The linear projection matrix $\mathbf{A}\in \mathbb{R}^{M\times N}$ is also known as the sensing matrix or measurement matrix in CS, which can be generated randomly \cite{wen2018deep,liang2020deep}.
The compressed codeword $\mathbf{s}$ is quantized into a bitstream vector $\mathbf{s}_q$ via an uniformed quantization operation $\mathsf{Qua}(\cdot)$ \cite{Jang2019quantization,guo2022overview} with the quantization bits $B$, i.e., $\mathbf{s}_q=\mathsf{Qua}(\mathbf{s};B)\in \{0,1\}^{B \times M}$. 

Then, the bitstream codeword is fedback to BS via a perfect uplink channel.\footnote{We assume perfect feedback link where no error occurs on the codeword $\mathbf{s}_q$.} It is subsequently restored into the real-valued codeword vector $\mathbf{s}_d$ via a dequantization operation $\mathsf{Deq}(\cdot)$, i.e., $\mathbf{s}_d=\mathsf{Deq}(\mathbf{s}_q)$, where $\mathsf{Deq}(\cdot)$ is the inverse of  $\mathsf{Qua}(\cdot)$. 
A decoder NN is deployed at BS to decompress and reconstruct the original CSI matrix $\mathbf{H}$ from $\mathbf{s}_d$. Denote the decoder function as $\mathsf{Dec}(\cdot)$ and the recovered CSI matrix is expressed as:

\begin{equation}
    \widehat{{\mathbf H}}= \mathsf{Dec}({\mathbf s}_d;\boldsymbol{\omega}),
\end{equation}
where $\boldsymbol{\omega}$ denotes the parameters of the decoder NN.

\subsection{Task Characteristics}
\label{Scenario characteristics}
The CSI module is affected by various factors in different scenarios and tasks. In addition to system requirements and limitations on the model (reconstruction accuracy, feedback overhead, NN complexity, etc.), factors affecting CSI feature distribution are also included, such as scene complexity, mobile terminal speed, sampling intensity, bandwidth, and antenna distribution. Among these, scene complexity is particularly important, as each scenario has unique environmental characteristics, such as the distribution (density or height) of buildings, trees, and other features that depict the UE's surrounding scattering environment.
In practice, the BS collects the downlink CSI by region, where the CSI maps sampled within a local subregion exhibit high spatial correlation and reflect the scattering environment around UE in the scene\cite{QuaDRiGa,cost2100channel}. To solve the problem of excessive communication cost caused by the collection of high-resolution downlink CSI at BS, the uplink CSI is used instead of downlink CSI \cite{Song2021Machine,Utschick2022Learning} in NAS searching and training to obtain the optimal network, and then a small amount of downlink CSI samples are used to fine-tune the network parameters.

The multi-path frequency response channel vector ${\mathbf h}_n$ on the $n^{\rm th}$ subcarrier can be formulated as \cite{Utschick2022Learning}:
\vspace{-0.2cm}
\begin{equation}
    \mathbf{h}_{n} =\sum_{\ell=1}^{L}  \alpha_{\ell}\mathrm{e}^{j(\theta_{\ell}-2\pi f_{n}\tau_{\ell})} \mathbf{a}_{\mathrm t}^{\mathrm H}\left(\phi_{\ell}\right),
\label{eq:h-n}
\end{equation}
where $L$ and $f_{n}$ denote the propagation path number and the $n^{\rm th}$ subcarrier frequency with $\alpha_{\ell}$, $\theta_{\ell}$, $\tau_{\ell}$ and $\phi_{\ell}$ stand for the complex attenuation amplitude, random phase shift, delay, azimuth angles of departure (AoDs) associated with the $\ell^{\rm th}$ path, respectively. $\mathbf{a}_{\mathrm t}(\phi_{\ell})$ stands for the antenna array response vectors at the BS, and the ULA antenna array response vectors can be given as \cite{wen2014channel-estimation}:
\begin{equation}
    \mathbf{a}_{\mathrm t}\left(\phi_{\ell}\right) =\frac{1}{N_{\mathrm t}} \left[1, \mathrm{e}^{-j \varpi_{n} \sin (\phi_{\ell})},..., \mathrm{e}^{-j \varpi_{n}({N_{\mathrm t}}-1) \sin (\phi_{\ell})}\right]^{\mathrm T},
 \label{eq:alpha}
\end{equation}
in which $\varpi_{n}=2\pi d f_{n}/c$
with $c$ and $d$ denoting the speed of light and the distance between antenna elements, respectively.

\begin{figure*}
    \centering
    \subfigure[COST2100 Indoor and Outdoor scenarios\label{fig:PSE-cost2100}]{
			\includegraphics[width=0.45\linewidth]{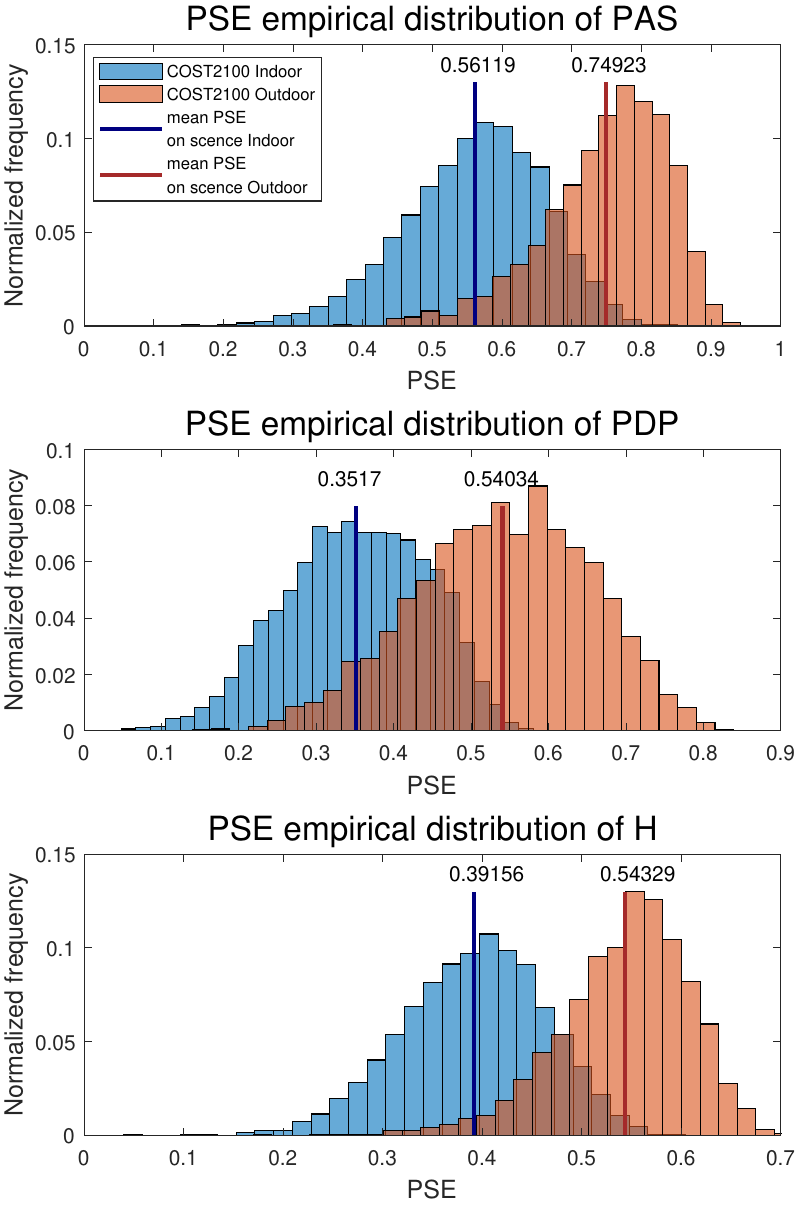}}%
 \subfigure[QuaDRiGa Scenario 1 (Park) and Scenario 2 (Commercial district)\label{fig:PSE-quadriga}]{
			\includegraphics[width=0.45\linewidth]{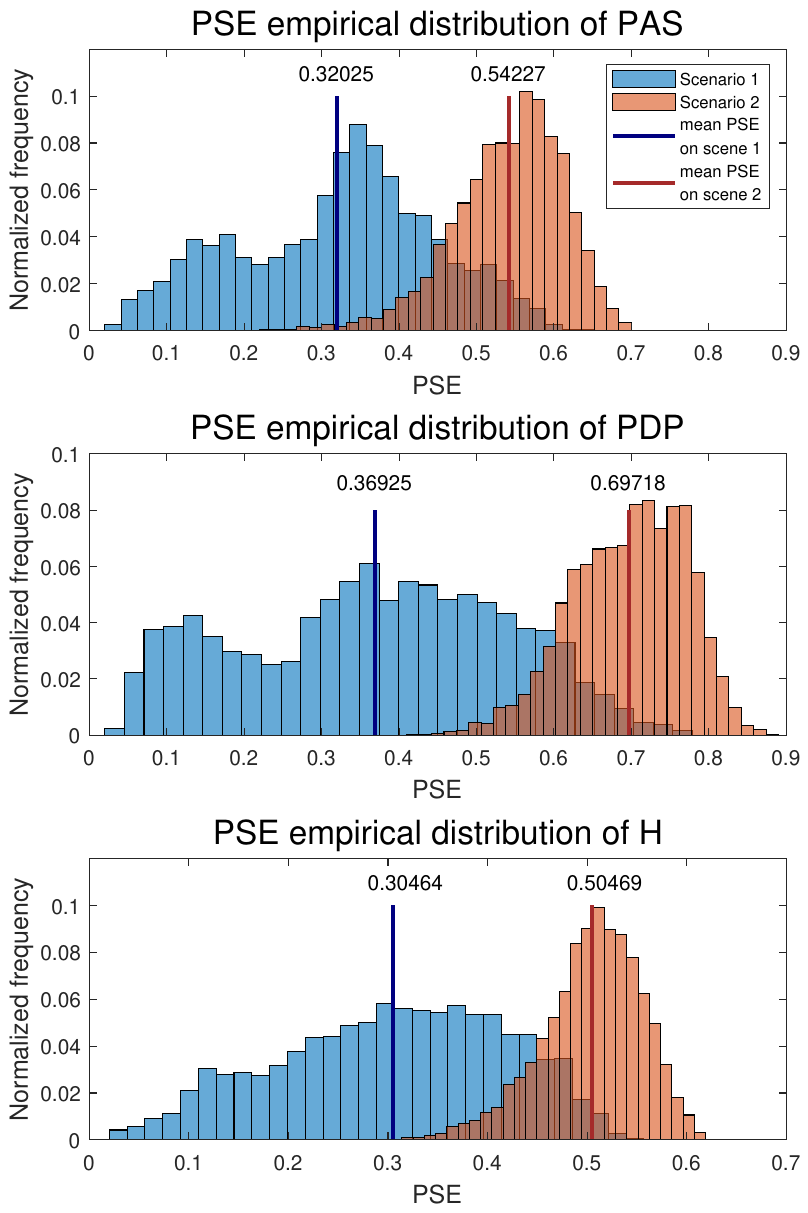}}%
    \caption{PSE empirical distribution for PAS, PDP and $\mathbf{H}$ on COST2100 dataset (left) and QuaDRiGa simulating dataset (right).}
    \label{fig:PSE}\vspace{-0.7cm}
\end{figure*}

Equation (\ref{eq:h-n}) describes the relationship between scenario complexity and CSI feature sparsity. For example, a rich scattering environment, such as a commercial district, is reflected in a dense and complex CSI with a wide angular and delay range. In contrast, the CSI map of an open scattering environment, such as a park, is sparse, with the feature centrally distributed. To describe the 1-D feature distributions, we also use the power delay profile (PDP) and power angular spectrum (PAS), which can be calculated as:
\vspace{-0.15cm}
\begin{eqnarray}
\mathrm{PAS}(\mathbf{H})&=& \frac{1}{N_{\mathrm t}} \left [ \left \| \mathsf{col}_k(\mathbf{H} ) \right \|_2^2  \right ]_{k=1}^{N_{\mathrm t}}, \nonumber\\
\mathrm{PDP}(\mathbf{H})&=& \frac{1}{N_{\mathrm c}}\left [ \left \| \mathsf{row}_k(\mathbf{H} ) \right \|_2^2  \right ]_{k=1}^{N_{\mathrm c}},
\label{equ-PAS-PDP}
\end{eqnarray}
where $\mathbf{H}$, $\mathsf{col}_k(\cdot)$ and $\mathsf{row}_k(\cdot)$ denote the CSI matrix, and $k^{\rm th}$ column and row of matrix, respectively. $||\cdot ||_2$ stands for the Euclidean norm. The distribution of channels in the angular and delay domain (i.e., $\mathrm {PAS}$ and $\mathrm {PDP}$) has certain statistical characteristics under specific scenarios.
To further quantify the compressibility of the CSI matrix for different scenarios, we introduce power spectral entropy ($\mathrm {PSE}$)\cite{powell1979spectral} as the compressibility metric. For a given $K$-length vector $\mathbf{v}=[v_1,v_2,...,v_K]$, the $\mathrm {PSE}$ of $\mathbf{v}$ can be defined as
\begin{equation}
    {\mathrm {PSE}}(\mathbf{v})=-\frac{1}{\log_{2}{K} } \sum_{i=1}^{K} p(v_i)\log_{2}{p(v_i)},
\end{equation}
where $p(v_i)={|v_i|^2}/({\textstyle \sum_{i=1}^{K}} |v_i|^2 )$ stands for the probability of the component $v_i$. The $\mathrm {PSE}$ value is normalized to $[0,1]$. From the perspective of information theory, low information entropy means that the information source is relatively certain and the corresponding amount of information is small. Therefore, a low $\mathrm {PSE}$ value indicates that CSI has a high degree of compressibility.
Figure \ref{fig:PSE} shows the $\mathrm {PSE}$ distribution of $\mathrm {PAS}$, $\mathrm {PDP}$, and $\mathbf {H}$, where two scenarios are marked in blue and orange, respectively. The COST2100 dataset in Figure \ref{fig:PSE-cost2100} is the same as in \cite{wen2018deep}. In Figure \ref{fig:PSE-quadriga}, QuaDRiGa software\footnote{QuaDRiGa software [Online] Available: \url{https://quadriga-channel-model.de/}.} \cite{QuaDRiGa} was used to simulate two scenarios, namely, a park scene (with empty buildings) and a business district scene (with dense buildings) in a typical urban community. The statistic $\mathrm{PSE}$ is calculated with 100,000 CSI samples in each scenario, which is adequate to reflect the scene characteristics. The orange-marked scenario (COST2100 Outdoor or QuaDRiGa simulated scenario 2) is more complex than the blue-marked scenario (COST2100 Indoor or QuaDRiGa simulated scenario 1), as reflected in higher $\mathrm {PSE}$ values and low compressibility in CSI samples. For the COST2100 Indoor/Outdoor scenario, the $\mathrm {PSE}$ value of $\mathrm {PAS}$ is higher than that of $\mathrm {PDP}$, indicating higher compressibility in the delay domain, while the two scenarios of QuaDRiGa show the opposite situation.
Scenario 1 of QuaDRiGa has a wider distribution of $\mathrm {PSE}$ compared to scenario 2 because $\mathrm {PSE}$ is more sensitive to sparse features in the CSI.

\subsection{Task-specific NN Architecture}
Manual design of NN structures is resource-intensive, making it challenging to construct customized NN structures for specific tasks in practice. As a result, researchers often resort to using a general NN structure for all scenarios or tasks, which may result in lower performance for each specific task. This is due to the limited generalization of the NN structure, making it difficult to design a general NN structure that performs well across all scenarios or tasks.
The following is the analysis from three perspectives:
\paragraph{Relationship between NN capacity and the CSI information entropy}
For homogeneous tasks like CSI feedback in different scenarios, the required NN capacity varies with the data distributions and CSI information entropy. For example, the COST2100 Outdoor scene is more complex than the Indoor scene, as shown in Figure \ref{fig:PSE-cost2100}. Table \ref{Tab:NN-archi-performance} shows that the CSI feedback network with the best performance in the Outdoor scene, MRFNet \cite{hu2021mrfnet}, is more complex than the one with the best performance in the Indoor scene, STNet \cite{mourya2022spatially}. Compared to STNet, MRFNet has a wider (more feature maps) and deeper (more layers) NN architecture, allowing it to remember more feature information during reconstruction.

\paragraph{Balancing NN performance and transplantability}
While a network structure can be portable and embeddable, NN structures with good transplantability, such as Transformer, LSTM, and Inception, are relatively simple unit structures. Small and general NNs have high portability but perform poorly due to limited representation ability, as they can only extract shallow features. In contrast, complex and dedicated task-oriented NNs have high performance but low transplantability due to their sophisticated architecture. For instance, transplanting the whole GoogLeNet framework to some object identification tasks may be disappointing, and it is common practice to transplant only GoogLeNet's Inception units. Therefore, a universal network to handle CSI feedback for all scenarios comes at the cost of sacrificing some model performance for each specific scenario.

\paragraph{Solution space and solving efficiency}
The NN architecture defines the solution space for a task, which expands as the architecture becomes more complex. However, a larger solution space is not always better for different scenarios. Complex scenes require a large solution space due to the difficult task, while simple scenes only require an appropriately sized solution space. In a large solution space, the solution of gradient descent can converge to an unsatisfactory local optimal point due to the inappropriate initial point, which leads to network degradation. This degradation phenomenon is supported by the results in Table \ref{Tab:NN-archi-performance}: the complex MRFNet \cite{hu2021mrfnet} failed to compete with the simpler STNet \cite{mourya2022spatially} on the Indoor task in terms of both performance and NN complexity.

To achieve optimal performance for specific tasks and scenarios, a universal NN would require comprehensive consideration of all task characteristics. Designing a universal NN structure that can accommodate all tasks is impractical. Therefore, it is necessary to customize NN structures according to specific scenarios, which requires a convenient and efficient scheme of NN structure customization.
\begin{figure*}[t]
    \centering
    \includegraphics[width=0.95\linewidth]{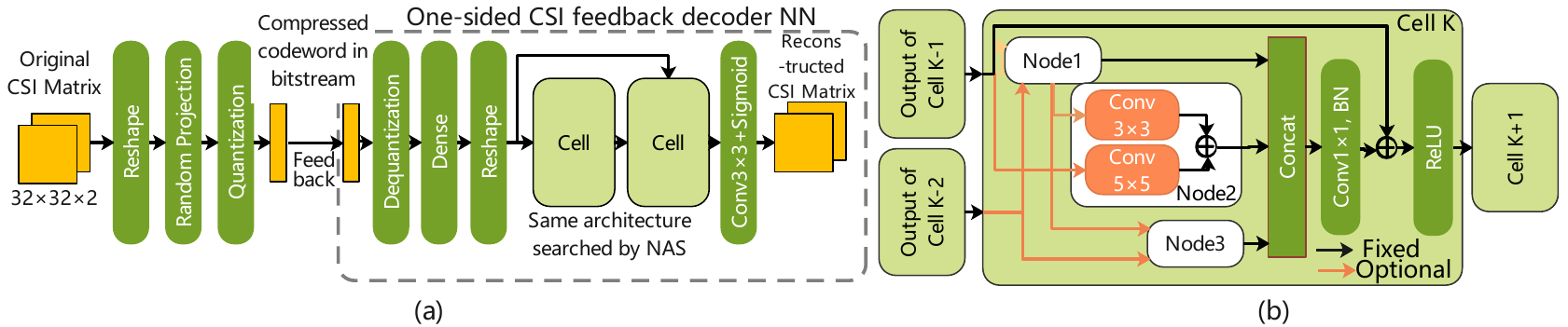}
    \setlength{\abovecaptionskip}{-0mm}
    \setlength{\belowcaptionskip}{-0cm}
    \caption{(a) Overall Auto-CsiNet architecture, where the reconstructed cell is searched using NAS. (b) The cell structure featuring a skip connection.}
    \label{fig:Auto-CsiNet}
    \vspace{-6mm}
\end{figure*}
\section{Framework of NAS-based automatic NN design}
\label{section:framework}

Manual design of a dedicated CSI feedback network for specific scenarios is a challenging task.
On one hand, people often assemble network operators or modules with limited expert experience and knowledge and through a random combination of ``trial and error'', which is inefficient, resource-intensive and experience-dependent.
On the other hand, scenario customized NN structure design can further obtain higher performance, while scene customization is difficult to achieve by manual design due to the high design cost.
In addition, there is no theoretical research on the relationship between scene knowledge and the CSI feedback NN structure, which can not guide the artificial design of scene-customized NN structure.

To address the challenges associated with manual design, AutoML is a promising solution for building DL systems without human assistance, thereby reducing the dependence on human experiential knowledge and broadening the application domain of DL. One essential component of AutoML is NAS, which aims to optimize the hyperparameters of NN structure in DL. NAS has been widely studied and successfully applied in various AI fields, such as pattern recognition, image segmentation, semantic analysis, etc.

Given the efficiency of NAS, we propose using NAS to replace manually designing NN architectures and harness the full potential of machine learning in designing CSI feedback networks. This automatic CSI feedback NN architecture generation scheme enables convenient and efficient design of the optimal NN structure for a specific scenario, and can be regarded as a black box in operation, whose input is the CSI dataset of a specific scene and output is the optimal NN structure of the scene. Inside the black box, the search loop is conducted based on three core elements in NAS: search space, search strategy, and evaluation strategy. A search space, consisting of candidate NN structures, is established. An NN is selected from the search space based on the search strategy. The model is subsequently evaluated, and its performance on the validation set is used to update the search strategy. The loop continues until a stopping condition is met, and the optimal NN structure is achieved. Through this data-driven manner, the implicit scene knowledge hidden in the CSI feature distribution can be explored and then be utilized in the scenario-customized NN architecture design process.

NAS is known to demand significant computing power and hardware resources, resulting in a high implementation threshold. However, by selecting appropriate and efficient NAS technologies and controlling the difficulty of the search, the automatic generation process for the CSI feedback NN structure in this paper consumes fewer resources and allows for cost control, making it easy to implement in actual production. Next, the NAS scheme, which has been adopted in our automatic generation process, is described from the perspective of the three elements in NAS. Each element is detailed in the following subsection.

\vspace{-0.3cm}
\subsection{Cell-based Search Space}
The search space defines the structural paradigm that NAS can explore, i.e., the solution space. If the solution space is too large, NAS can become overburdened, which may be difficult to support by current computing power devices. Cell-based networks, which are built by stacking several repetitive cells, are a smaller-scale solution space that significantly reduces search costs compared to the global search space, which allows arbitrary connections between ordered nodes and has the largest scale.
For example, the NASNet \cite{zoph2018learning} method to search the CIFAR-10 classification network requires 800 GPUs to run 28 days in the global search space, while the cost is reduced to 500 GPUs and 4 days in the unit-based search.
Additionally, most CSI feedback NN architectures that have been manually designed are cell-based (as depicted in Table \ref{Tab:NN-archi-performance}), which supports the selection of a cell-based search space as a reasonable and effective approach.

\begin{table}[t]
\centering
\setlength{\abovecaptionskip}{0cm}
    \setlength{\belowcaptionskip}{-0.2cm}
    \renewcommand\arraystretch{0.93}
	\caption{\label{Tab:Requirment} Configuration presetting.}
\begin{threeparttable}
\resizebox{0.49\textwidth}{!}{

\begin{tabular}{cl}
\toprule
    \multirow{4}{*}{\begin{tabular}[c]{@{}c@{}}Overall architecture \\ setting\end{tabular}} & CR: $\gamma$ \\& number of RefineNet cells: $N_{\mathrm{cell}}$   \\                                                    & number of cell's inner nodes: $N$    \\
    & cell's width (channel number): $c$ \\ \midrule
\multirow{7}{*}{NAS setting}                                                             & dataset partition ratio: $\mathbf{p}= [p_{\alpha}, p_{\omega},1-p_{\alpha}-p_{\omega}]$ \\
& warm up epoches: $ E_{\mathrm{warm\,up}} $ \\
& search epoches: $E_{\mathrm{search}}$        \\& optional operation set: $\mathcal{O}=\{o_m\}_{m=1}^{|\mathcal{O}|}$  \\
& architecture weight optimizer and learning rate: $\ell_{\alpha}$\\    & NN parameter optimizer and learning rate: $\ell_{\omega}$  \\
& architecture weight decay rate: $\lambda $ \\\midrule
\multirow{3}{*}{Evaluation setting} & evaluation training epochs: $E_{\mathrm{train}}$\\
& start record epochs: $E_{\mathrm{start\,record}}$\tnote{1}\\
& maximum number of records: $M_{\mathrm{record}}$\\
\bottomrule
\end{tabular}}
\begin{tablenotes}
	\footnotesize
	\item{1}: $E_{\mathrm{start\,record}}<E_{\mathrm{search}}$.
\end{tablenotes}

\end{threeparttable}
\vspace{-5mm}
\end{table}

\begin{figure*}[t]
    \centering
    \setlength{\abovecaptionskip}{-0mm}
    \includegraphics[width=0.95\textwidth]{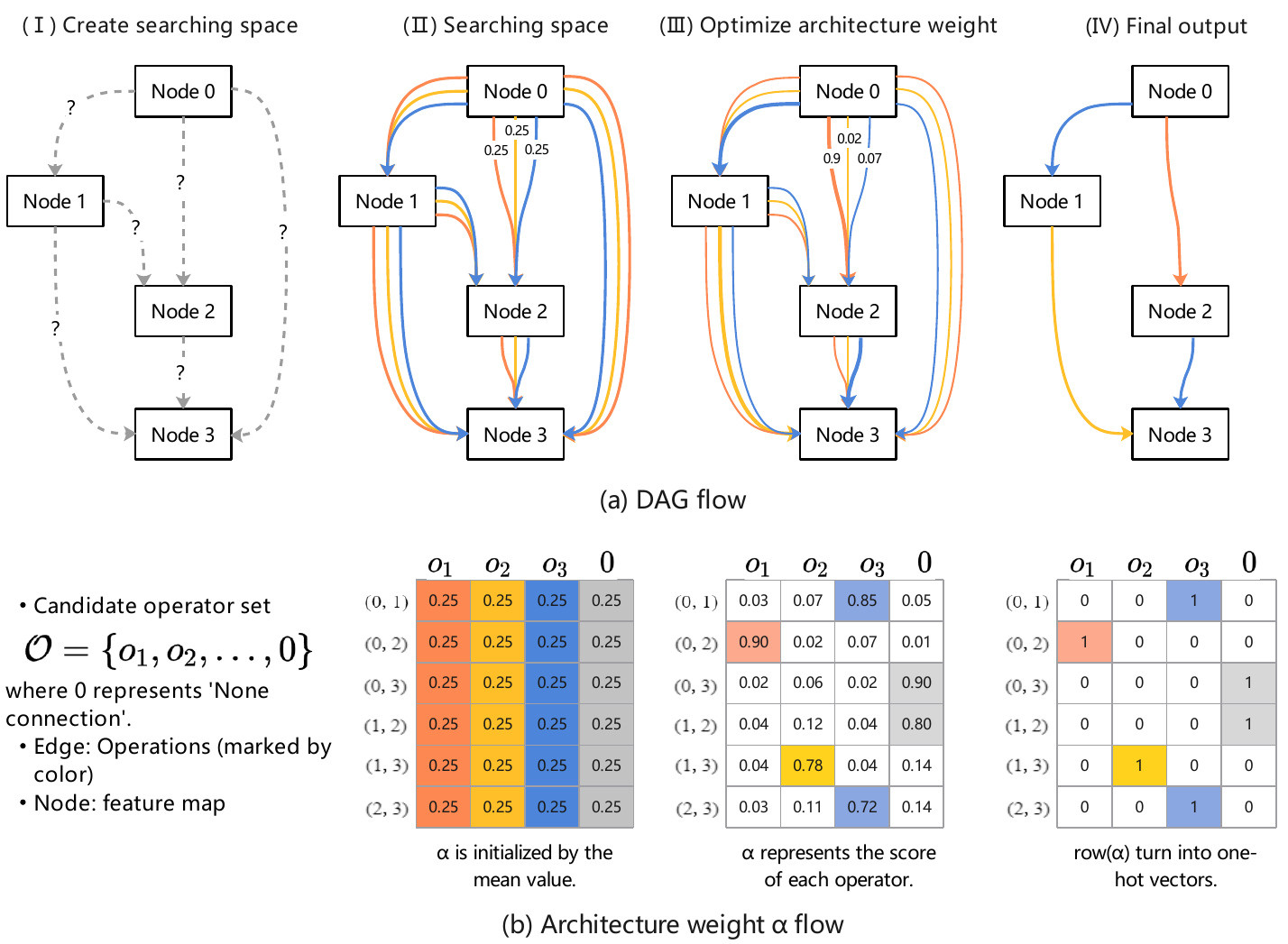}
    \caption{Pipeline of Gradient Descent-based searching strategy: DARTS\cite{DARTS}.}
    \label{fig:pipeline-DARTs}
    \vspace{-5mm}
\end{figure*}
\subsubsection{Overall Architecture of Auto-CsiNet}
Figure \ref{fig:Auto-CsiNet} depicts (a) the cell-based overall architecture of our proposed Auto-CsiNet, and (b) the cell's inner structure.
Auto-CsiNet is designed based on CS-CsiNet \cite{wen2018deep} by modifying the RefineNet cell searched by NAS. CSI dimension reduction in the encoder is achieved through random projection. The compression ratio (CS) represents the ratio of the compressed codeword's dimension to the original CSI dimension. The cell-based search space is created, and the two refinement cells share the same architecture (Figure \ref{fig:Auto-CsiNet}(b)). The cell $k$ has two external input nodes connected to the output of the former cell $k-1$ and the cell $k-2$. Each cell has several inner nodes (3 nodes in Figure \ref{fig:Auto-CsiNet}(b)), and each node contains two branches, each with a selectable operation and input assignment. The two branches are added to get the node output, and the output of all internal nodes is concatenated. The channel value of the concatenated feature node is restored through the $1\times 1$ convolutional layer. To increase the flow of feature information in the cell and avoid network degradation, we add residual connections, consistent with most manual-designed works (as depicted in Table \ref{Tab:NN-archi-performance}). The output of the current unit is obtained after passing through the ReLU layer. The first part in Table \ref{Tab:Requirment} lists the structural parameters that need to be configured in advance when setting up Auto-CsiNet (which cannot be changed in the automatic generation stage), including CR, the number of repetitive cells, the cell's inner nodes and width (channel value).

\subsubsection{Space Complexity}
We discuss the space scale, i.e., the number of the candidate architectures of Auto-CsiNet. Given a cell with $N$ inner nodes, equal-sized to a $2N$-layer global-based structure, and a candidate operation set $\mathcal{O}$, the cell-based space scale is:
\begin{equation}
   \prod_{k=0}^{N-1} \left | \mathcal{O}  \right | ^2\times \mathrm{C}_{k+2}^{2},
   \label{equ:cel-space}
\end{equation}
where $\mathrm{C}$ stands for composite. $\left | \mathcal{O}  \right | ^2$ indicates that each node contains two branches (two operations), and $\mathrm{C}_{k+2}^{2}$ represents the number of combinations of node $k$ connected to the preceding nodes. The equal-sized global-based space scale is $\left | \mathcal{O}  \right |^{2N}\times 2^{\mathrm{C}_{2N}^{2}}$, where $2^{\mathrm{C}_{2N}^{2}}$ represents two possibilities between any two layers, connected or not connected. From the expressions, the cell-based space scale is much smaller than the global-based one. For instance, $\left | \mathcal{O}  \right |=8, N=5$ in our later experiments, the cell-based scale is 2.89e12, while the equal-sized global-based scale is 3.78e22. By selecting the cell-based search space, the space scale is significantly reduced, which effectively controls the difficulty and cost of searching.

\subsection{Gradient Decent Search Strategy}
The search strategy defines how to find the optimal network structure, which is a hyperparameter optimization problem of the NN structure, tuned according to the performance of NN observed in the validation set. The most efficient, convenient, and accessible search strategy thus far is based on the gradient descent (GD) method \cite{DARTS,PC-DARTS}, which can significantly reduce the computing power and time resource consumption compared with other NAS methods.For example, in search of the CIFAR              -10 classification network, Large-scale ensemble \cite{Large-Scale-Evolution}, an evolution algorithm NAS method, requires 250 GPUs to run 10 days, while DARTS only takes one GPU and 0.4 days.

\subsubsection{Pipeline of DARTS}
\begin{figure*}[t]
    \centering
    \includegraphics[width=0.95\textwidth]{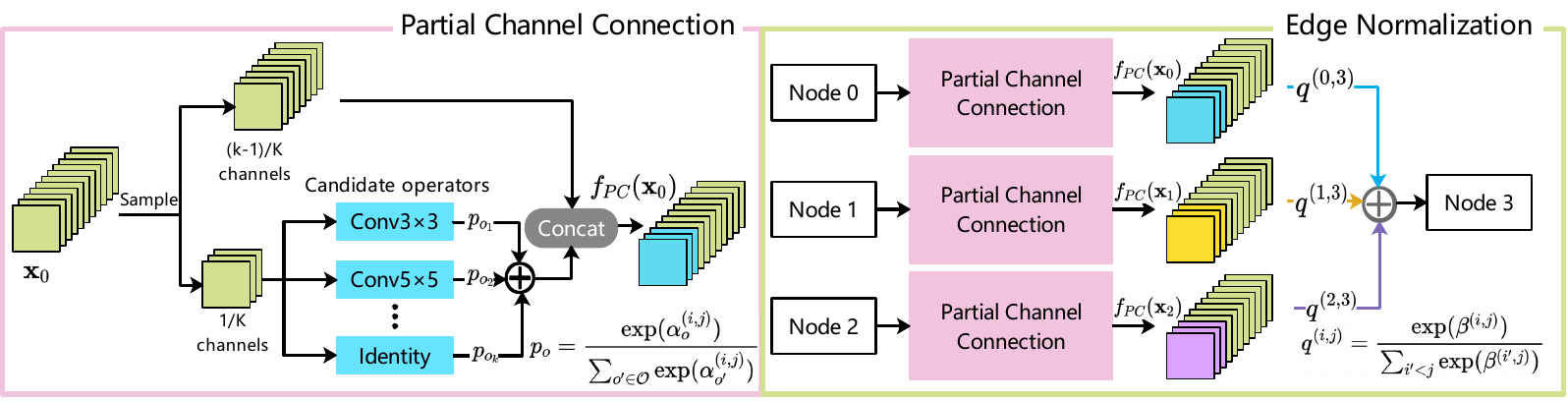}
    \caption{Partial-Channel-Connection and the Edge-Normalization operations in PC-DARTS \cite{PC-DARTS}. Using the topology example shown in Figure \ref{fig:pipeline-DARTs}, the process of information propagation to node 3 ($j=3$) is illustrated as an example.}
    \label{fig:PC-DARTs}
    \vspace{-0.5cm}
\end{figure*}
Different from the other iterative search strategies based on discrete search space (the number of candidate network structures is countable), the core idea of GD-based algorithm is to make the discrete optimization problem (strategy function) continuous, which the GD-based algorithm can optimize.
The most typical GD-based approach is DARTS \cite{DARTS}, which treats the NAS process as a black box problem and maps the original discrete search space to the continuous space through a particular relaxation operation.

The cell's topological structure is a directed acyclic graph (DAG) consisting of an ordered sequence of N nodes (as depicted in Figure \ref{fig:pipeline-DARTs}(a)). Each node $\textbf{x}^{(i)}$ stands for the feature map in convolutional networks, and each  directed edge $(i,j)$ is associated with some operation $o^{(i,j)}$ that transforms $\textbf{x}^{(i)}$ to $\textbf{x}^{(j)}$.
Each intermediate node is computed based on all of its predecessors:
\begin{equation}
    \textbf{x}^{(j)}=\sum_{i<j}o^{(i,j)}(\textbf{x}^{(i)}).
\end{equation}
Let $\mathcal{O}=\{o_1,o_2,...,0\}$ be the set of candidate operations, e.g., convolution, max pooling, $\verb|zero|$, where a special $\verb|zero|$ operation is included to indicate a lack of connection between two nodes.
Then, the discrete search space is relaxed into continuous by placing a mixture of candidate operations on each edge, that is, relaxing the categorical choice of a particular operation to a softmax over all possible operations:
\begin{equation}
    \bar{o} ^{(i,j)}=\sum_{o\in \mathcal{O} }\frac{\exp (\alpha _o^{(i,j)})}{\sum_{o'\in \mathcal{O} }\exp (\alpha _{o'}^{(i,j)})}o ^{(i,j)},
\end{equation}
where the operation mixing weights for edge $(i,j)$ are parameterized by a vector $\{\alpha_o^{(i,j)}\}_{o\in \mathcal{O}}$ of dimension $|\mathcal{O}|$. In this way, the task of the architecture search is reduced to learning a set of continuous variables $\{\alpha_o^{(i,j)}\}_{o\in \mathcal{O},i<j}$, which we refer to as the architecture weight (encoded as) and depicted in Figure \ref{fig:pipeline-DARTs}(b). The architecture weights in the figure have been normalized.

Figure \ref{fig:pipeline-DARTs} presents the pipeline of DARTS \cite{DARTS}, and the detailed steps are as follows: (I) Construct the search space and determine the candidate operators. (II) Build the SuperNet corresponding to the search space and relax the search space by placing the softmax mixture of candidate operations. The assigned architecture weights $\{\mathbf{\alpha}_o^{(i,j)}\}_{o\in \mathcal{O},i<j}$ are initialized with the mean value. (III) SuperNet optimization: jointly optimize the architecture weights and the NN parameters (e.g., kernel parameters of the convolutional layer) by solving a bi-level optimization. The thickness of the edge represents the architecture weight. (IV)
The operator with the largest weight (the coarsest edge) is retained to combine into the final output NN structure, i.e., replace each mixed operation $\bar{o}^{(i,j)}$ with the most likely operation $o ^{(i,j)} =\mathrm{argmax} _{o\in \mathcal{O} } \alpha _o^{(i,j)}$\footnote{Since in cell-based space, each node has an in-degree of 2, the first two with the highest probability will be retained when selecting the operators.}.

\subsubsection{Advanced Version PC-DARTS}
The NAS method adopted in our automatic CSI feedback generation scheme is PC-DARTS \cite{PC-DARTS}, an advanced version of DARTS, which involves the Partial-Channel-Connection and the Edge-Normalization operations to improve the search efficiency of DARTS further. As depicted in Figure \ref{fig:PC-DARTs}, the high searching speed mainly comes from the Partial-Channel-Connection operation, where all candidate operators only apply to $1/K$ part of the feature maps (marked by colors), and the rest is directly concatenated to the operating-acted part. This operation also addresses the problem of excessive memory consumption in the SuperNet training of DARTS \cite{DARTS} (SuperNet is $|\mathcal{O}|$ times more complex than normal networks).

The hyper-parameter $K$ in the proportion $1/K$ of selected channels can be adjusted and represents the multiple by which PC-DARTS outperforms DARTS in search speed. In our experiments, we set $K=4$ as in \cite{PC-DARTS}. However, the negative effect of Partial-Channel-Connection, that is, the convergence instability of the architecture weight caused by random selection of channels in each round, may lead to undesired final search results. Edge-Normalization is adopted to reduce the fluctuation effect by introducing another set of normalized weights $\{\beta^{(i,j)}\}$ that act on the edges:
\begin{equation}
    \mathbf{x}_j=\sum_{i<j}\frac{\exp (\beta^{(i,j)})}{\sum_{i'<j }\exp (\beta ^{(i',j)})} f_{\mathrm{PC}}(\mathbf{x}_i),
\end{equation}
where $f_{\mathrm{PC}}$ denotes the function of Partial-Channel-Connection, involving the calculation of $\{\alpha_o^{(i,j)}\}_{o\in \mathcal{O}}$. Therefore, the connectivity of edge $(i,j)$ is determined by both $\{\mathbf{\alpha}_o^{(i,j)}\}_{o\in \mathcal{O}}$ and $\beta^{(i,j)}$, for which the normalized coefficients are multiplied together, i.e., $\frac{\exp (\beta^{(i,j)})}{\sum_{i'<j }\exp (\beta ^{(i',j)})} \times \frac{\exp (\alpha _o^{(i,j)})}{\sum_{o'\in \mathcal{O} }\exp (\alpha _{o'}^{(i,j)})}$. Then, the operations (edges) are selected by finding the large edge weights
as in DARTS.
Therefore, $\boldsymbol{\beta}$ can dilute the fluctuation effect of $\boldsymbol{\alpha}$ that acts on the unfixed feature maps.
For convenience, we denote $\boldsymbol{\alpha}$ as the whole architecture weights (i.e., $\{\alpha_o^{(i,j)};\beta^{(i,j)}\}_{o\in \mathcal{O},i<j}$).
Thanks to the Partial-Channel-Connection and Edge-Normalization operations, the advanced PS-DARTS can finish the search on the c-10 dataset within several hours, significantly reducing the search cost.

\begin{algorithm}[t]
	\caption{Automatic generation for scenario-customized NN architecture}
     \label{alg:Automatic-generation}
	\LinesNumbered 
	\KwIn{The scenario-specific CSI dataset: $ \mathcal{D} $;
        the required configuration presettings in Table \ref{Tab:Requirment}.}
	\KwOut{The optimal scenario-customized NN architecture: $ A^* $.}
	$\mathcal{D}_{\mathrm{train},{\alpha}}$, $\mathcal{D}_{\mathrm{train},{\omega}}$, $\mathcal{D}_{\mathrm{test}}$ ← Divide $\mathcal{D}$ randomly according to the preset ratio $\mathbf{p}$\; 
     Construct the SuperNet with $\{\gamma, N_{\mathrm{cell}}, N, c  \}$\;
     Initialize the architecture weights $\mathbf{\alpha}$ with the mean value and initialize the NN parameters $\mathbf{\omega}$ randomly\;
     $i$ ← 0\;
	\While{$i<E_{\mathrm{search}}$ or $not \;converged$}{

		\eIf{$i<E_{\mathrm{warm\,up}}$}{
			Fix the architecture weights $\boldsymbol{\alpha}$\;
		}{
			Update the architecture weights $\boldsymbol{\alpha}$ by descending $\bigtriangledown _{\alpha } \{\mathcal{L}_{\mathrm{MSE}}(\boldsymbol{\omega},\boldsymbol{\alpha} ;\mathcal{D}_{\mathrm{train},{\alpha}})
            +\lambda \left \| \boldsymbol{\alpha} \right \| _2^2\}$\;
		}
          Update the NN parameters $\omega$ by descending $\bigtriangledown _{\omega } \mathcal{L}_{\mathrm{MSE}}(\boldsymbol{\omega},\boldsymbol{\alpha} ;\mathcal{D}_{\mathrm{train},{\omega}})$\;
          $i$ ← $i$+1\;
	}
    Derive the final architecture based on the learned $\boldsymbol{\alpha}^*$: $ A^* $.

\end{algorithm}
\subsection{Weight-sharing Evaluation Strategy}
The evaluation strategy (evaluation forecast) is required to predict the approximate NN performance and accelerate the evaluation process. Among them, weight-sharing is a core driving force of the GD-based approach to achieve efficient search. In DARTS \cite{DARTS} and PC-DARTS \cite{PC-DARTS}, the one-shot weight-sharing enables the search and evaluation to be conducted in parallel. Instead of training each architecture separately as the manual design process, weight sharing builds a SuperNet that assembles all the architectures as its submodels. In Figure \ref{fig:pipeline-DARTs}(a),
the search space can be viewed as the SuperNet (Super-DAG) with all candidate operators and connections, from which the sub-network (sub-DAG) is sampled and evaluated.
Once the SuperNet has been trained, sub-network can be directly evaluated by inheriting weights from the SuperNet, thus saving the considerable cost of training each architecture from scratch.

\subsubsection{Bi-level Optimization in SuperNet}

The SuperNet contains edge/architecture weights $\boldsymbol{\alpha}$ and the corresponding operator's parameters $\omega$, alternating optimized in each training epoch to realize the bi-level optimization.

The scenario-specific CSI training dataset $\mathcal{D}_{\mathrm{train}}$ is divided into two parts, $\mathcal{D}_{\mathrm{train},\omega}$ and $\mathcal{D}_{\mathrm{train},\alpha}$. The mean square error (MSE) loss is adopted in CSI feedback NN training:
\begin{eqnarray}
\mathcal{L}_{\mathrm{MSE}}(\boldsymbol{\alpha}, \boldsymbol{\omega};\mathcal{D}) =  \frac{1}{|\mathcal{D}| }\sum_{\mathbf{H}\in \mathcal{D}}\left \| \mathsf{SuperNet}(\boldsymbol{\alpha}, \boldsymbol{\omega};\mathbf{s}_d)-\mathbf{H} \right \|_2^2,
\end{eqnarray}
where $\mathbf{s}_d$ denotes the compressed code after the dequantization given in \eqref{equ3}.
Then, the bi-level optimization problem with $\boldsymbol{\alpha}$ as the upper-level variable and $\boldsymbol{\omega}$ as the lower-level variable:
\begin{eqnarray}
\min_{\boldsymbol{\alpha}}  &&\mathcal{L}_{\mathrm{MSE}}(\boldsymbol{\alpha},\boldsymbol{\omega}^*(\boldsymbol{\alpha});\mathcal{D}_{\mathrm{train},\alpha}) +  \lambda \left \| \boldsymbol{\alpha} \right \| _2^2 \\ \nonumber
\mathrm{s.t.} &&\boldsymbol{\omega}^*(\boldsymbol{\alpha})=  \mathrm{argmin} _{\boldsymbol{\omega}} \ \mathcal{L}_{\mathrm{MSE}}(\boldsymbol{\alpha},\boldsymbol{\omega};\mathcal{D}_{\mathrm{train},\omega}),
\end{eqnarray}
where $\left \| \boldsymbol{\alpha} \right \| _2^2$ is the $L_2$ regular term with respect of $\boldsymbol{\alpha}$, promoting the high score (probability) to appear in a certain operator, which is convenient to determine the structure of the final search. $\lambda$ is the weight decay rate. The goal for architecture search is to find the optimal $\boldsymbol{\alpha}^*$ that minimizes the loss on $\mathcal{D}_{\mathrm{train},\alpha}$, which can be regarded as minimizing the validation loss. The optimal NN parameters $\boldsymbol{\omega}^*(\boldsymbol{\alpha})$ are obtained by minimizing the loss on $\mathcal{D}_{\mathrm{train},\omega}$ (regarded as the training loss). Since the second order hessian matrix is involved in calculating the gradient of $\boldsymbol{\alpha}$ associated with $\boldsymbol{\omega}$, leading tremendous calculation burden, we adopt the first order approximation, i.e., $\bigtriangledown _{\alpha }\mathcal{L}_{\mathrm{MSE}}(\boldsymbol{\alpha},\boldsymbol{\omega}^*(\boldsymbol{\alpha});\mathcal{D}_{\mathrm{train},{\alpha}}) \approx  \bigtriangledown _{\alpha }\mathcal{L}_{\mathrm{MSE}}(\boldsymbol{\alpha},\boldsymbol{\omega};\mathcal{D}_{\mathrm{train},{\alpha}})$, where the $\boldsymbol{\alpha}$ and $\boldsymbol{\omega}$ are updated alternately in each epoch.

Algorithm \ref{alg:Automatic-generation} depicts our proposed automatic generation scheme for scenario-customized CSI feedback NN architecture, including the alternative bi-level optimization process.
For lower-level optimizations, operator parameters are updated while architecture weights are fixed. For higher-level optimizations, the opposite is true. Each sub-network can directly inherit the parameters of the selected operator, and its evaluation value can directly influence the architecture weight of the corresponding operators. In addition, the warm-up stage is used for NN parameters to grow up fully, where the architecture weights are not updated. Otherwise, the SuperNet would prefer selecting the parameter-free operations, such as Identity or pooling, which converge faster than those complex ones.

\begin{algorithm}[t]
	\caption{Automatic generation scheme with early stopping}
     \label{alg:Early-stopping}
	\LinesNumbered 
	\KwIn{The scenario-specific CSI dataset: $ \mathcal{D} $;\\
        the required configuration presettings in Table \ref{Tab:Requirment};\\
        }
	\KwOut{The optimal scenario-customized NN architecture: $ A^* $.}
	Conduct line 1-3 of Algorithm \ref{alg:Automatic-generation}\;
     $i$ ← 0; $val_{\mathrm{nmse}}$ ← 0 [dB]; $\mathcal{A}$ ← $\{\}$\;

	\While{$i<E_{\mathrm{search}}$ or $not \;converged$}{\tcp{Searching stage}
		Conduct line 6-12 of Algorithm \ref{alg:Automatic-generation}\;
        Calculate $\mathrm{NMSE}(\sf{SuperNet};\mathcal{D}_{\mathrm{test}})$ by \eqref{equ:NMSE}\;
        \uIf{$\mathrm{NMSE}(\mathsf{SuperNet};\mathcal{D}_{\mathrm{test}}) \le val_{\mathrm{nmse}}$}{
        Update $val_{\mathrm{nmse}}$ with $\mathrm{NMSE}(\mathsf{SuperNet};\mathcal{D}_{\mathrm{test}})$\;
        Derive the architecture $A$ based on the learned $\boldsymbol{\alpha}$\;
        \lIf{$i>E_{\mathrm{start\,record}}$ and $\left | \mathcal{A}  \right | \le M_{\mathrm{record}}$ and $A\notin \mathcal{A}$}{
        $\mathcal{A}$ ← $\{A\} \cup \mathcal{A}$}
        }
	}
     \ForEach{$A\in \mathcal{A}$}{\tcp{Evaluating stage}Training architecture $A$ with dataset $\mathcal{D}_{\mathrm{train},{\alpha}}\cup \mathcal{D}_{\mathrm{train},{\omega}}$ from scratch\;
     Evaluate $\mathrm{NMSE}(A;\mathcal{D}_{\mathrm{test}})$;
     }
    Derive the final optimal architecture $A^*=\underset{A\in \mathcal{A} }{\mathrm{argmin} }\; {\mathrm{NMSE}(A;\mathcal{D}_{\mathrm{test}})}$.

\end{algorithm}

\subsubsection{Early Stopping and Elastic Selection Criteria}
The measurement of reconstruction accuracy in the CSI feedback is normalized \rm{MSE} (NMSE):
\begin{equation}
    \mathrm{NMSE}(A;\mathcal{D} )=\frac{1}{\left | \mathcal{D} \right | } \sum_{\mathbf{H}\in \mathcal{D}}\frac{\left \| \mathbf{H}-A(\mathbf{s}_d) \right \|_2^2 }{\left \| \mathbf{H} \right \|_2^2},
    \label{equ:NMSE}
\end{equation}
where $\mathbf{s}_d$ denotes the dequantized compressed code given in \eqref{equ3}, and $A$ represents a trained CSI feedback NN architecture.

However, the hard criteria condition specified in Algorithm 1, which selects the output of the $E_{\mathrm{search}}$-th epoch as the optimal solution, is not flexible enough and may result in suboptimal structures due to convergence fluctuations. Furthermore, our experiments have shown that increasing the number of search rounds may not continuously improve the sub-network performance and may actually lead to performance degradation due to excessive searching. Therefore, we introduce the early stopping and elastic selection mechanisms in Algorithm \ref{alg:Early-stopping} to prevent over-searching.
Since the suitable searching epochs number $E_{\mathrm{search}}$ is also scenario-dependent and cannot simply be fixed, we record the structure of the output when SuperNet validates well, i.e., $A\in \mathcal{A}$, where $\mathcal{A}$ is the set of candidate
architectures. These elements are retrained during the evaluation phase, and the final optimal architecture, $A^*$, is selected from $\mathcal{A}$ based on its evaluation performance.

\section{Simulation Results and Discussions}\label{section:Simulation-Results}
In this section, we present details on the CSI dataset simulation and NN training. We then compare Auto-CsiNet, which is the architecture generated by the proposed automatic generation scheme for scenario-customized CSI feedback architecture, with manually designed works. We also investigate the characteristics of the searched architecture with respect to the searching time and scenario.

\subsection{CSI Simulation and Training Configuration}
\subsubsection{Scenario Setup}
For performance evaluation, we use four datasets. The first two QuaDRiGa datasets were simulated using QuaDRiGa software \cite{QuaDRiGa}, and the last two COST2100 datasets were simulated using COST2100 simulation software \cite{cost2100channel} and published by \cite{wen2018deep}. The simulation details are presented in Table \ref{tab:QuaDRiGaSetting}, while the scenario characteristics are discussed in Section \ref{Scenario characteristics}.

\begin{table*}[t]
    \setlength{\abovecaptionskip}{0cm}
    \setlength{\belowcaptionskip}{-0.2cm}
    \centering
    \caption{\label{tab:QuaDRiGaSetting}Basic parameter setting in simulation.}
    \renewcommand\arraystretch{0.9}
	\resizebox{0.99\textwidth}{!}{	
    \begin{tabular}{c|cc|cc}\toprule

Simulation platform                                                                           & \multicolumn{2}{c|}{QuaDRiGa}                                   & \multicolumn{2}{c}{COST2100}                                \\\hline
Dataset                                                                                  & \begin{tabular}[c]{@{}c@{}}QuaDRiGa Scene 1\\ (Park)\end{tabular} & \begin{tabular}[c]{@{}c@{}}QuaDRiGa Scene 2\\ (Commercial district)\end{tabular} & COST2100 Indoor             & COST2100 Outdoor              \\\hline
Antenna setting                                                                                & \multicolumn{4}{c}{32 ULA antennas at BS, single antenna at UE}                                                              \\\hline
Operating system                                                                               & \multicolumn{2}{c|}{FDD-OFDM system with 512 subcarriers}       & \multicolumn{2}{c}{FDD-OFDM system with 1024 subcarriers}   \\\hline
Center frequency                                                                               & \multicolumn{2}{c|}{2.655GHz}                                   & \multicolumn{1}{c}{5.3GHz}                      & 300MHz                        \\\hline
Bandwidth                                                                                      & \multicolumn{2}{c|}{20MHz}                                      & \multicolumn{2}{c}{20MHz}                                   \\\hline
Scenarios                                                                                      & \multicolumn{1}{c}{3GPP-38.901-UMi-LOS}    & 3GPP-38.901-UMi-NLOS                  & \multicolumn{2}{c}{\multirow{3}{*}{Not given in \cite{wen2018deep}}} \\\cline{1-3}
Space correlation distant                                                                      & \multicolumn{2}{c|}{20m}                                        & \multicolumn{2}{c}{}                                        \\\cline{1-3}
Scattering clusters number                                                                     & \multicolumn{1}{c}{5}                      & 20                                    & \multicolumn{2}{c}{}                                        \\\hline
Sampling range                                                                                 & \multicolumn{2}{c|}{$20m\times 20m$}                            & \multicolumn{1}{c}{$20m\times 20m$}             & $400m\times 400m$             \\\hline
\begin{tabular}[c]{@{}c@{}}Position of the sampling \\ area center relative to BS\end{tabular} & \multicolumn{1}{c}{(10,-70) m }             & (50,0) m                               & \multicolumn{2}{c}{(0,0) m}                                  \\\hline
Sampling number                                                                                & \multicolumn{2}{c|}{100,000} & \multicolumn{2}{c}{150,000}      \\\hline
CSI Pretreatment                                                                               & \multicolumn{4}{c}{2D-DFT, sparse clipping (reserve nonzero 32 rows), normalized to real values in the range $[0,1]$}\\
\bottomrule                                  \end{tabular}}
\end{table*}

\begin{table*}[t]
    \setlength{\abovecaptionskip}{0cm}
    \setlength{\belowcaptionskip}{-0.2cm}
    \renewcommand\arraystretch{0.9}
    \centering
    \caption{\label{tab:Training-Setting}Training settings for the searching stage and evaluation stage.}

	\resizebox{0.99\textwidth}{!}{	
    \begin{tabular}{l|l|l}\toprule
              & Searching stage & Evaluation stage                                                                     \\\hline
Dataset       &
\begin{tabular}[c]{@{}l@{}}QuaDRiGa: $|\mathcal{D}_{\mathrm{train},{\omega}}|:|\mathcal{D}_{\mathrm{train},{\alpha}}|:|\mathcal{D}_{\mathrm{test}}|=[0.5,0.45,0.05]$\\
COST2100: $|\mathcal{D}_{\mathrm{train},{\omega}}|:|\mathcal{D}_{\mathrm{train},{\alpha}}|:|\mathcal{D}_{\mathrm{test}}|=[10,3,2]$\\
Quantization bits $B=8$, CR=1/4\end{tabular}
&
\begin{tabular}[c]{@{}l@{}}QuaDRiGa: $|\mathcal{D}_{\mathrm{train}}|:|\mathcal{D}_{\mathrm{val}}|:|\mathcal{D}_{\mathrm{test}}|=[0.85,0.10,0.05]$\\
COST2100: $|\mathcal{D}_{\mathrm{train}}|:|\mathcal{D}_{\mathrm{val}}|:|\mathcal{D}_{\mathrm{test}}|=[10,3,2]$\\
Quantization bits $B=8$, CR=1/4 (by default)\end{tabular}
 \\\hline
Loss          &  \begin{tabular}[c]{@{}l@{}}
Architecture weight optimize: $L_2$ regularization with $\lambda$ of 3e-4; \\
NN parameter optimize: $\mathrm{MSE}$\end{tabular}
&   NN parameter optimize: $\mathrm{MSE}$    \\   \hline
Learning rate &       \begin{tabular}[c]{@{}l@{}}Architecture weight: Adam optimize, learning rate=6e-4;\\ NN parameter: Adam optimize, Exponential decay \\ (initial learning rate=3e-4, decay rate=0.97)\end{tabular} & \begin{tabular}[c]{@{}l@{}}NN parameter: Adam optimize, Exponential decay \\ (initial learning rate=1e-3, decay rate=0.97)\end{tabular}  \\\hline
Epochs        &   $E_{\mathrm{warm\,up}}=20$, $E_{\mathrm{search}}=400$, $E_{\mathrm{start\,record}}=20$                                                                                                                                           & $E_{\mathrm{training}}=1200$                                                                                     \\\hline
Cell settings &    \begin{tabular}[c]{@{}l@{}}Inner node number=\{1,2,3,4,5,6\}\\ Channel number = 8\\ Without residual connection (between the input node to the output)\end{tabular}                                   & \begin{tabular}[c]{@{}l@{}}Channel number = 7\\ With residual connection\end{tabular} \\\hline
GPU days      &    1 (Tesla V100) GPU $\times$ 0.4-1.0 days                     &    1 (Tesla V100) GPU $\times$ 0.1 days  \\\bottomrule
\end{tabular}}
\vspace{-6mm}
\end{table*}

\subsubsection{Training Configuration of Auto-CsiNet}
Table \ref{tab:Training-Setting} shows the training details for both the searching stage (SuperNet training) and the evaluation stage (sub-network training). In the searching stage, a search space based on cells is created, where the number of internal nodes is set as 1-6, and the cell's channel value is set to 8. The channel value or width of a cell represents the channel value of its internal nodes.
The candidate operators are as follows: { $\verb|zero|$, $\verb|skip_connection|$, $\verb|sep_conv3x3|$, $\verb|dil_conv3x3|$, $\verb|dil_conv5x5|$, $\verb|conv3x3|$, $\verb|conv1x5_5x1|$, $\verb|conv1x9_9x1|$},  which mean none connection, identity mapping, depth-separable convolution and dilated convolution, respectively.
Note that the cell structure used for building the SuperNet does not include a residual connection between the input and output nodes. This is because the candidate operators cannot be fully trained with such a connection. However, the residual connection is added to the searched cell structure during the sub-network evaluation. The rest of the configurations in the cell follow the standard cell-based methods proposed in \cite{DARTS,PC-DARTS}.

After a warm-up period of 20 epochs, the formal search stage is initiated, as described in Algorithm \ref{alg:Automatic-generation} and \ref{alg:Early-stopping}. Following 400 epochs of searching, a sub-optimal candidate architecture set $\mathcal{A}$ with the scale $M_{\mathrm{record}}=20$ is obtained, greatly reducing the scale of the cell-based space from 2.89e12 (when inner node number is 5) to 20. The sub-networks' training and evaluation follow the standard setup used in CsiNet \cite{wen2018deep}. Notably, the sub-network generated through the search process need not be consistent with the SuperNet's width. In this experiment, the sub-network channel value is set to 7 by default to compare with the manual design work \cite{lu2020multi}. The experiment is performed using a single Tesla V100 GPU, and the search stage takes 0.4-1.0 days (depending on the cell's node configuration), while the evaluation stage takes 1.0-2.0 days (which requires about a dozen sub-networks to be trained, each taking approximately 0.1 days). Furthermore, if GPU memory is sufficient, the search and evaluation stages can be conducted in parallel to further reduce the time cost.

\subsection{Experiments and Analysis}


\subsubsection{Effectiveness of Auto-CsiNet}

\begin{figure*}[t]
    \centering
    \setlength{\abovecaptionskip}{-2mm}
    \includegraphics[width=0.95\linewidth]{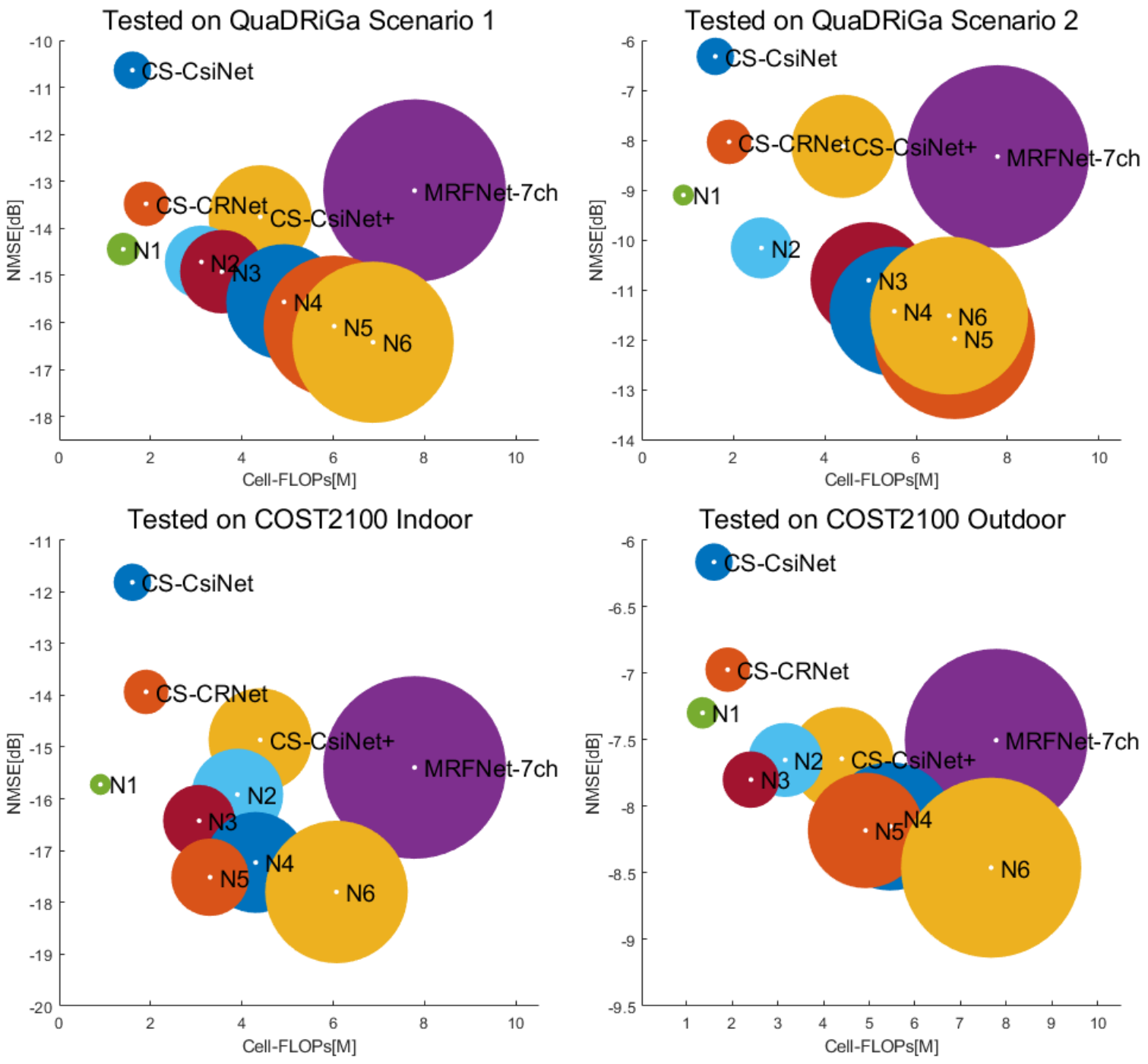}
    \caption{Balance of cell complexity and NN performance of Auto-CsiNet series (N1-N6) in QuaDRiGa scene 1/2 and COST2100 Indoor/Outdoor scenarios at CR = 1/4 and $B=8$.}
    \label{fig:efficiency}
    \vspace{-0.4cm}
\end{figure*}

\begin{table*}[t]
    \setlength{\tabcolsep}{3.5pt}
    \renewcommand\arraystretch{0.8}
    \caption{\label{Tab:NN-runtime} 
    NMSE performance and running time of Auto-CsiNet vs. manual-designed networks on COST2100 Indoor dataset with a quantization bit of $B=8$. Running time is measured at CR=1/4.}
    \centering
    \resizebox{0.9\textwidth}{!}{
    \begin{threeparttable}
    \begin{tabular}{cc|cccc|cccccc}
\toprule

& \multirow{2}{*}{CR}            & \multirow{2}{*}{\begin{tabular}[c]{@{}c@{}}CS-CsiNet \\ \cite{wen2018deep}\end{tabular}} & \multirow{2}{*}{\begin{tabular}[c]{@{}c@{}}CS-CRNet\\ \cite{lu2020multi}\end{tabular} } & \multirow{2}{*}
{\begin{tabular}[c]{@{}c@{}}CS-CsiNet+ \\ \cite{guo2020convolutional}\end{tabular}} & \multirow{2}{*}{\begin{tabular}[c]{@{}c@{}}CS-MRFNet \\ \cite{hu2021mrfnet}\end{tabular} }                     & \multicolumn{6}{c}{Auto-CsiNet}                                    \\
& &      &      &    &    & N1      & N2      & N3      & N4               & N5     &N6          \\ \midrule
\multirow{5}{*}{NMSE [dB]} &1/4                       & -11.99  & -13.34 & -13.90   &-14.13
& -14.56     & -15.31  & -15.86  & -16.36  & -17.09   & \textbf{-17.85} \\
& 1/8 & -7.95  & -9.80 & -9.98   &-9.27  & -9.52     & -10.07  & -10.32  & -10.79  & \textbf{-11.05}   & -10.91           \\
& 1/16 & -4.95  & -5.82 & -5.61   & -6.02 & -5.68     & -5.68  & -6.56  & \textbf{-6.79}  & -6.45   & -6.40           \\
& 1/32 & -3.03   & -3.71 & -3.31  & -3.76 & -4.00  & -4.17  & -4.32  & -4.52  & \textbf{-4.55}  & -4.47 \\
& 1/64 & -2.46   & -2.29 & -3.02  & -3.04 & -2.11  & -2.03  & -2.22  & -3.02  & -3.21   & \textbf{-3.30}    \\ \midrule

\multicolumn{2}{c|}{running time per sample [s]} &4.36e-02   & 4.62e-02   &4.36e-02   & 4.73e-02
& 4.60e-02 & 4.65e-02  & 4.67e-02 &5.04e-02 &4.97e-02& 4.86e-02   \\
\multicolumn{2}{c|}{Cell complxity in FLOPs [M]}&1.62& 1.91& 4.37& 7.78& 0.90& 3.91& 3.06& 4.30& 3.32& 6.07\\
\bottomrule
\end{tabular}
     \end{threeparttable}
    }
\end{table*}
\begin{table*}[t]
    \setlength{\tabcolsep}{3.5pt}
    \renewcommand\arraystretch{0.8}
    \caption{\label{Tab:quantized-CSI}
    NMSE performance of NNs at CR=1/4 on COST2100 Indoor dataset using uniform quantization with various quantization bits.}
    \centering
    \resizebox{0.8\textwidth}{!}{
    \begin{threeparttable}
    \begin{tabular}{c|cccc|cccccc}
\toprule

Quantization bits            & \multirow{2}{*}{\begin{tabular}[c]{@{}c@{}}CS-CsiNet \\ \cite{wen2018deep}\end{tabular}} & \multirow{2}{*}{\begin{tabular}[c]{@{}c@{}}CS-CRNet\\ \cite{lu2020multi}\end{tabular} } & \multirow{2}{*}
{\begin{tabular}[c]{@{}c@{}}CS-CsiNet+ \\ \cite{guo2020convolutional}\end{tabular}} & \multirow{2}{*}{\begin{tabular}[c]{@{}c@{}}CS-MRFNet \\ \cite{hu2021mrfnet}\end{tabular} }                     & \multicolumn{6}{c}{Auto-CsiNet}                                    \\
$B$ &      &      &    &    & N1      & N2      & N3      & N4               & N5     &N6          \\ \midrule
4 & -10.86 & -12.29   & -12.77 & -12.91
& -14.06    & -14.39 & -14.82 & -15.43 & \textbf{-16.62}   & -16.48 \\
8 & -11.99  & -13.34 & -13.90   &-14.13
& -14.56     & -15.31  & -15.86  & -16.36  & -17.09   & \textbf{-17.85}  \\
32 & -12.22  & -13.93 & -14.86   & -15.40
& -15.72 & -15.92 & -16.42 & -17.23 & -17.98 & \textbf{-18.21}       \\

\bottomrule
\end{tabular}
     \end{threeparttable}
    }
\end{table*}
There is a tradeoff between NN complexity and performance, and a practical cell structure can achieve a better balance between the two. To compare the tradeoff effectiveness between NAS-CsiNet and manual-designed networks, we compare the performance of Auto-CsiNet with that of CS-CsiNet \cite{wen2018deep}, CS-CsiNet+ \cite{guo2020convolutional}, CS-CRNet \cite{lu2020multi} and CS-MRFNet \cite{hu2021mrfnet}. For comparison, we use the RefineNet in CsiNet-M1 \cite{guo2020convolutional}, the CRBlock \cite{lu2020multi}, and the MRFBlock \cite{hu2021mrfnet} as the cell structure in CS-CsiNet+, CS-CRNet, and CS-MRFNet, respectively. The cell width is set to 7 for all networks. We conduct the search on four scenario datasets (QuaDRiGa Scene 1/2 and COST2100 Indoor/Outdoor) and obtain the corresponding Auto-CsiNet series (N1-N6) with a variety of internal nodes in the cell structure. The results are shown in Figure \ref{fig:efficiency}, where the different models are marked by different colors, and the radius represents the complexity of the cell structure. Note that the cell complexity is affected by the number of internal nodes and the operator selection. On the balance diagram for all scenarios, the NAS-CsiNet series are located at the lower left equilibrium points compared to the artificial-designed networks. For example, Auto-CsiNet-N1 surpasses CS-CsiNet or CS-CRNet in NMSE performance and is more lightweight. These results demonstrate that Auto-CsiNet can exceed the performance and efficiency of the manual-designed NN, thus verifying the effectiveness of Auto-CsiNet.

\begin{figure*}[t]
    \centering
    \subfigure[CRBlock]{\includegraphics[width=0.25\linewidth]{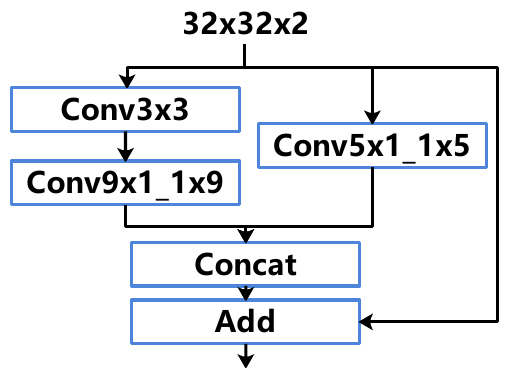}}
    \subfigure[DAG(CRBlock)]{\includegraphics[width=0.66\linewidth]{CRBlock.pdf}}
    \caption{CRBlock representation in the search space.}
    \label{fig:CRNet}
    \vspace{-0.4cm}
\end{figure*}
In addition, it is worth noting that the manual-designed CS-CRNet \cite{lu2020multi} can be represented in the search space, specifically the CRBlock in CS-CRNet is one of the candidate structures for NAS search with N3 configuration (as shown in Figure \ref{fig:CRNet}). Thus, as the optimal search result, Auto-CsiNet-N3 can achieve a better balance point than CS-CRNet within the same search space. This demonstrates that NAS can quickly search for the optimal solution in the search space with a reasonable and efficient optimization strategy. In this approach, PC-DARTS is utilized to employ GD-based optimization and the weight-sharing strategy to speed up the search. Compared with the inefficient random trial and error process of manually designed NN structure, this approach can traverse 4.72e6 [calculate by \eqref{equ:cel-space}] candidate structures in a limited time, and its valid and orderly traversal process also ensures its searching efficiency.

This advantage also explains why Auto-CsiNet can outperform manually designed networks such as MRFBlock \cite{hu2021mrfnet} when the cell structure is relatively complex. Large capacity networks experience a diminishing marginal effect, meaning that increasing model size yields diminishing returns, which is more pronounced in artificially designed networks. The size of the search space expands exponentially with the number of inner nodes of the cell, making it more difficult to manually design the network structure. Due to the limited capacity to traverse sub-networks manually in the large-scale search space, the probability of selecting the optimal solution is minimal. Therefore, powerful NAS is necessary to quickly traverse more sub-networks and reduce the diminishing marginal effect of complex units to a certain extent.

Additional results are presented in Tables \ref{Tab:NN-runtime} and \ref{Tab:quantized-CSI}, demonstrating a certain level of generalization with respect to the CR and quantization bits $B$. The proposed scheme proves effective even with minor deviations in the settings of CR and $B$ during the search and evaluation stages. Table \ref{Tab:NN-runtime} also details the network's practical running time per sample, tested on a Tesla V100 GPU with a batch size of 1. Notably, these results do not consistently align with the ranking of cell-FLOPs complexity. Factors such as data read, memory usage, and GPU occupancy affect this discrepancy. It is also attributed to the parallel operation of each branch in the multi-branch cell structure, where the network's runtime is determined by the longest-running branch, rather than the number of branches. Despite having higher cell complexity, Auto-CsiNet-N6 exhibits a shorter runtime than Auto-CsiNet-N4.
The calculation of cell FLOPs complexity involves the sum of the complexities of all operators/layers. 

\begin{figure*}[t]
    \centering
    \setlength{\abovecaptionskip}{0pt}
    \subfigure[ \label{fig:SE}]{
			\includegraphics[height=0.4\linewidth]{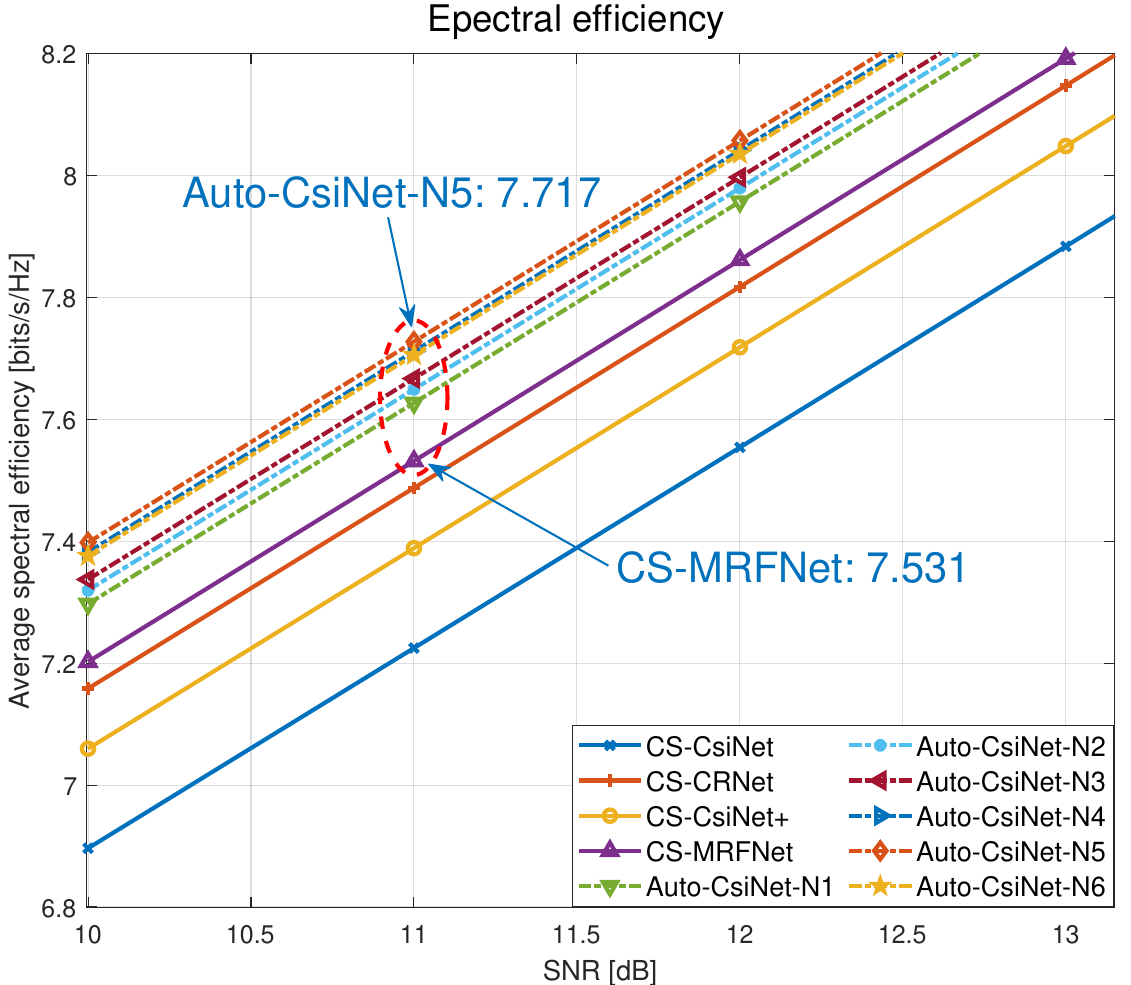}}%
 \subfigure[\label{fig:sample-size}]{
			\includegraphics[height=0.4\linewidth]{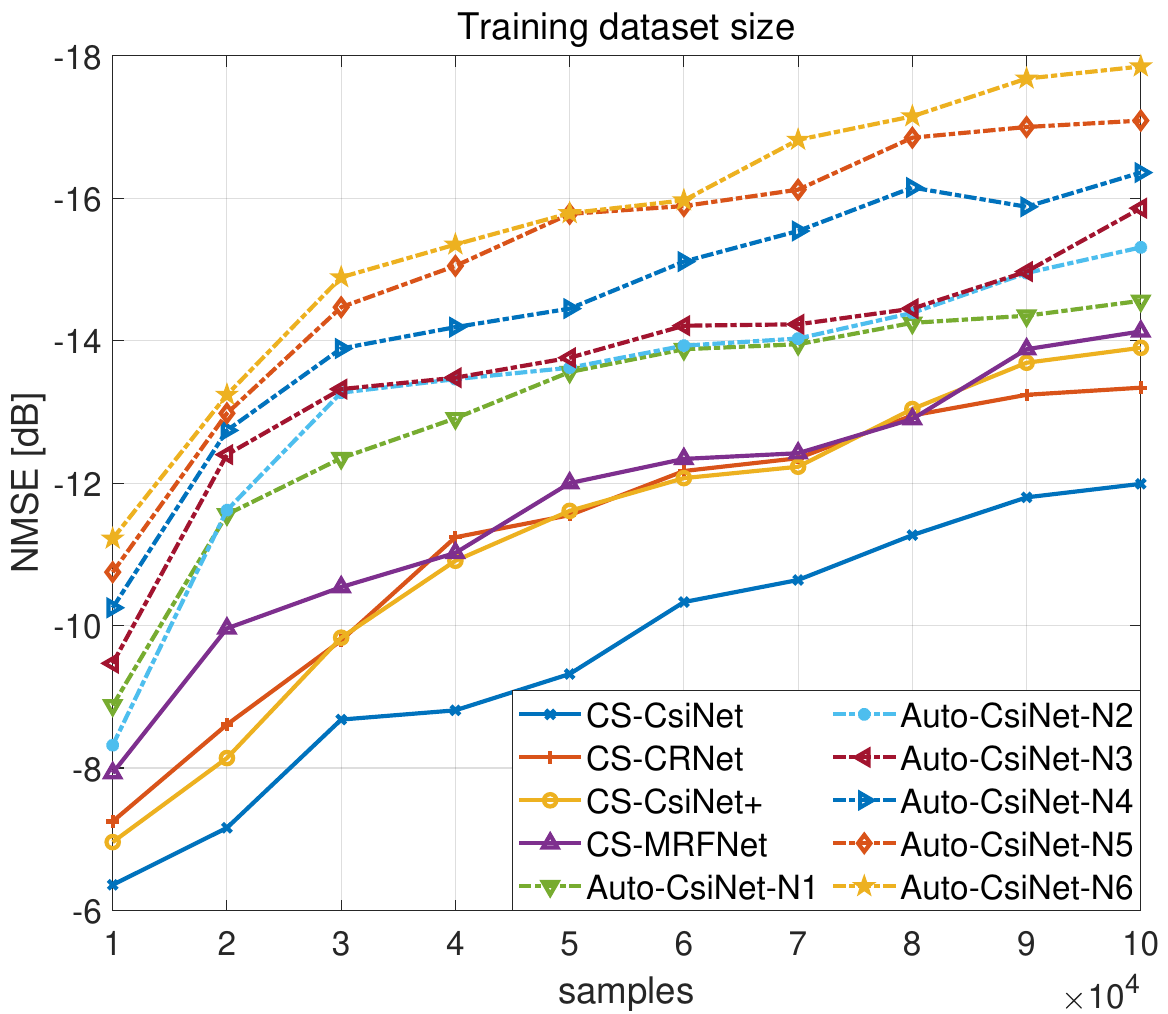}}%
    \caption{(a) Average spectral efficiency vs SNR at CR = 1/32 and $B=8$. (b) Training dataset size vs networks performance at CR=1/4 and $B=8$. Results are based on COST2100 Indoor scenario dataset.}
    \vspace{-4mm}
\end{figure*}

Figure \ref{fig:SE} showcases the average spectral efficiency of the MISO-OFDM system, which employs a zero-forcing precoding scheme. The spectral efficiency results are influenced by the performance of the decoder network used for CSI reconstruction. The NMSE performance of these decoder networks is detailed in Table \ref{Tab:NN-runtime}. We compare the state-of-the-art (SOTA) manually designed model, CS-MRFNet \cite{hu2021mrfnet}, with Auto-CsiNet-N5, highlighting their spectral efficiency values at an SNR of 11 dB. The performance gain ratio is approximately 2.47\% (a gain of 0.186 bits/s/Hz over 7.531 bits/s/Hz). This gain is attributed solely to modifications in the decoder network structure for CSI reconstruction. 

Furthermore, Figure \ref{fig:sample-size} illustrates the effect of training dataset size on network performance. The performance ranking is hardly affected by the training set size, ensuring the proposed scheme's effectiveness in search without large datasets. Although reducing training samples only yields approximate performance with reduced accuracy, it has negligible effects on decision-making in the search process, i.e., architecture weight optimization. This result also validates the strategy that a reduced training set can further accelerate model evaluation in NAS \cite{Klein2017Fast}.

\begin{figure}[t]
    \centering
    \setlength{\abovecaptionskip}{-2mm}
    \includegraphics[width=0.95\linewidth]{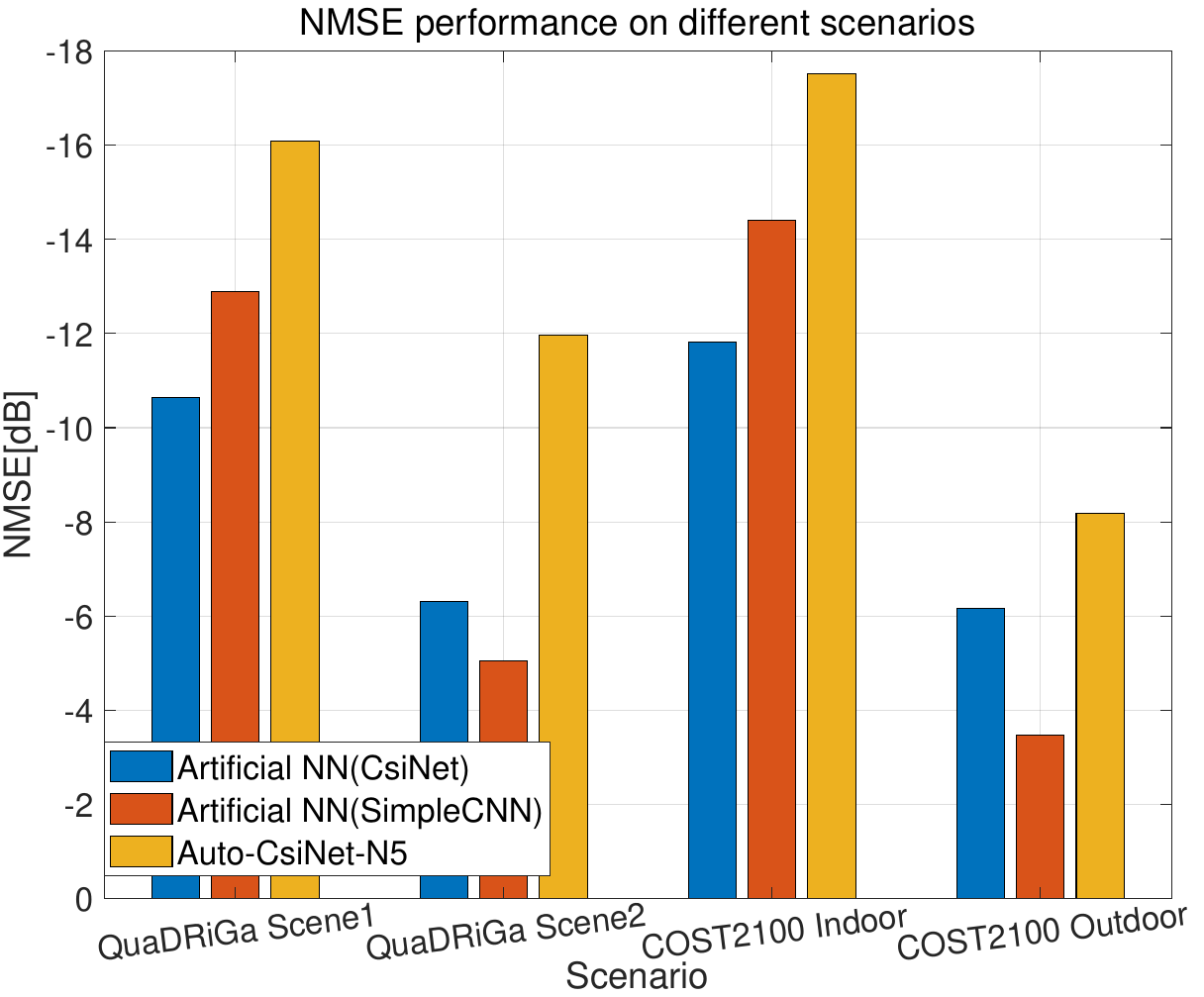}
    \caption{Comparison of manually-designed NN and scenario-customized Auto-CsiNet-N5 at CR = 1/4 and $B=8$.}
    \label{fig:gain-customize}
    \vspace{-0.4cm}
\end{figure}
\begin{figure*}[t]
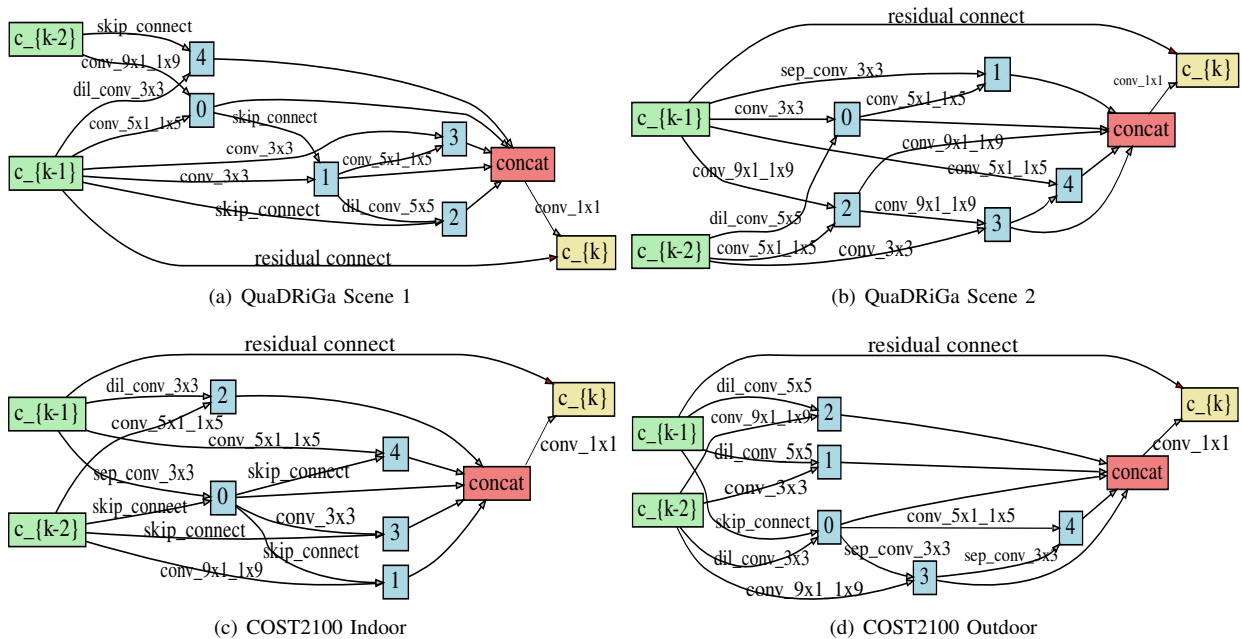

    \centering
    \subfigure[QuaDRiGa Scene 1\label{fig:archi-Qua1}]{\includegraphics[width=0.45\linewidth]{Node5_Qua11.pdf}}
	\subfigure[QuaDRiGa Scene 2\label{fig:archi-Qua2}]{\includegraphics[width=0.45\linewidth]{Node5_Qua21.pdf}}
	\subfigure[COST2100 Indoor\label{fig:archi-indoor}]{\includegraphics[width=0.45\linewidth]{Node5_indoor1.pdf}}
	\subfigure[COST2100 Outdoor\label{fig:archi-outdoor}]{\includegraphics[width=0.45\linewidth]{Node5_outdoor1.pdf}}
    \caption{Architectures of scenario-customized Auto-CsiNet-N5 at CR = 1/4 and $B=8$.}
    \label{fig:4type-Auto-CsiNet-N5}

\vspace{-0.6cm}
\end{figure*}
\subsubsection{Gain for Scenario Customization}
Due to the high cost of manual design work, it can often be challenging to customize NN structures to fit specific scenarios. The aim of Auto-CsiNet proposed in this paper is not only to replace manual design work with AutoML but also to obtain scenario-specific NN architectures conveniently by adopting low-cost NAS methods, thereby unleashing the maximum potential of NN for this task.

In Figure \ref{fig:gain-customize}, we compare Auto-CsiNet-N5 with two artificially-designed NNs, CS-CsiNet and CS-SimpleCNN. CS-SimpleCNN's decoder only contains a dense layer and a 3x3 convolutional layer, that is, CS-CsiNet removes two RefineNet units. We use Auto-CsiNet-N5 as an example of the Auto-CsiNet series to show the experimental comparison results, and the performance of other models in the Auto-Csinet-Nx series are shown in Figure \ref{fig:efficiency}. Auto-CsiNet-N5s are searched on the QuaDRiGa Scene 1/2 and COST2100 Indoor/Outdoor tasks, and the architectures are depicted in Figure \ref{fig:4type-Auto-CsiNet-N5}, while the manually-designed networks are general-purpose for all tasks.

The results demonstrate that a universal network is not adequate for all tasks, as the NN structure has limited generalization and may not show the same superiority for different tasks. For instance, while CS-CsiNet is more complex than CS-SimpleCNN, it cannot outperform CS-SimpleCNN on all tasks. CS-SimpleCNN performs better in the QuaDRiGa scene 1 and COST2100 Indoor scenario. In contrast, the customized-designed Auto-CsiNet-N5s achieve the highest performance for the specific scenario. The concept of scene-specific customization of Auto-CsiNet can be viewed as giving up the generalization of the NN structure in exchange for greater potential for performance improvement so that the NN can unleash its maximum potential for the given scenario.

\begin{figure*}[t]
    \centering
    \setlength{\abovecaptionskip}{-4mm}

    \includegraphics[width=0.99\linewidth]{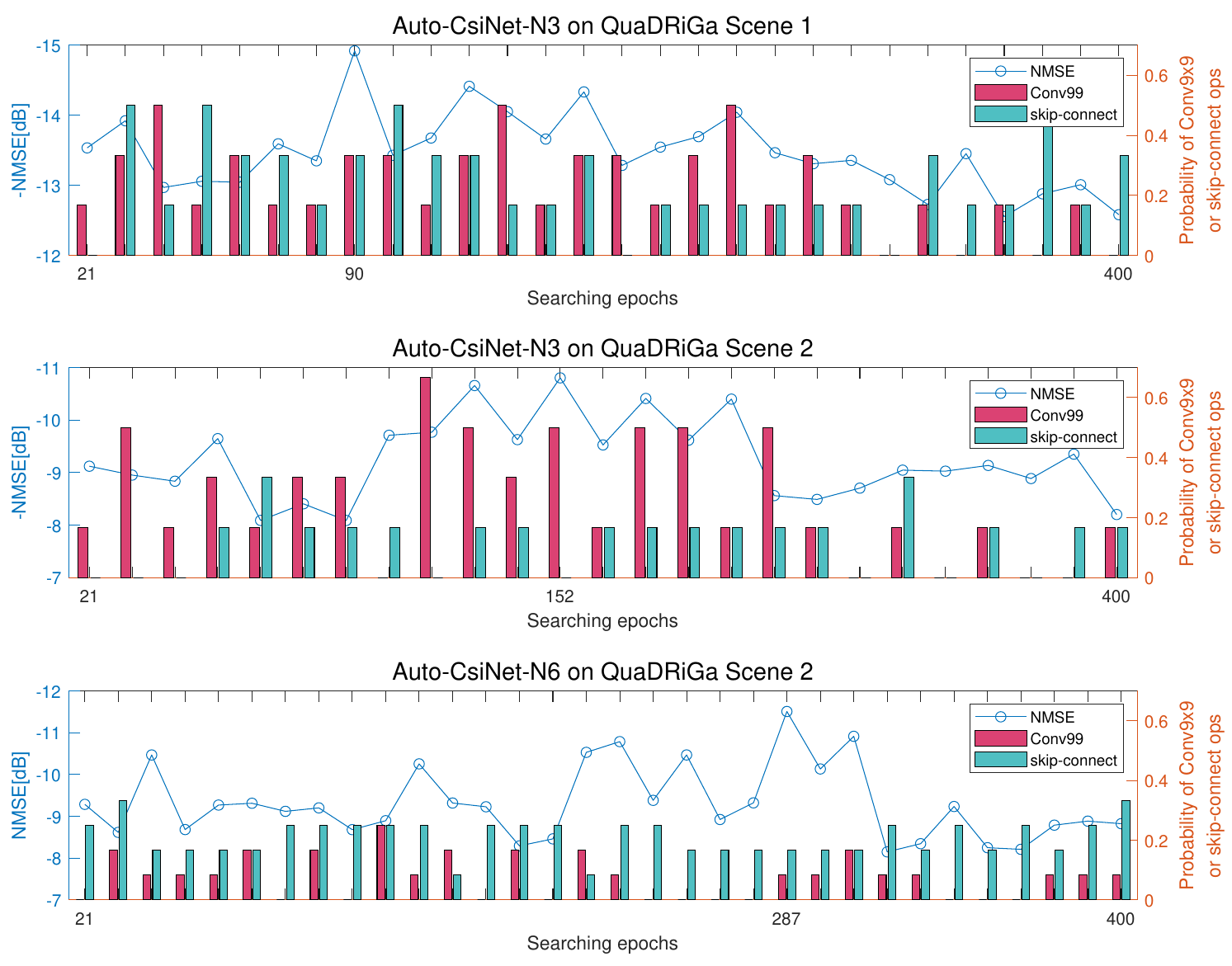}
    \caption{Sub-network architecture characteristic and NMSE performance during the search.}
    \label{fig:architecture characteristic}
    \vspace{-0.6cm}
\end{figure*}

\subsubsection{Architecture Characteristic Analysis}
Figure \ref{fig:architecture characteristic} illustrates the effect of the scenario on the architecture by displaying the search results of Auto-CsiNet-N3 on QuaDRiGa Scene 1 and Scene 2 (first two subfigures). It also compares the cell configuration effect by examining Auto-CsiNet-N3 and Auto-CsiNet-N6 searched on QuaDRiGa Scene 2 (last two subfigures).
In each subfigure, the bar represents the proportion of the simplest operator ($\verb|skip_connection|$) or the most complex operator ($\verb|conv1x9_9x1|$) within a cell, and the line depicts the the NMSE performance of the sub-network at each period in the search process.

In the first two subfigures of Figure \ref{fig:architecture characteristic}, it is observed that complex operators, such as $\verb|conv1x9_9x1|$ with a high parameter number, are more likely to appear in the search structure of complex scenes with high PSE. On the contrary, simple operators like $\verb|skip_connection|$ are more likely to appear in the search results of simple scenes.
Comparing Auto-CsiNet-N3 and Auto-CsiNet-N6 searched on a same scenario dataset, the probability of the complex convolution is higher than the parameter-free operation in Auto-CsiNet-N3, indicating that operator $\verb|conv1x9_9x1|$ can bring more gain to the sub-network than operator $\verb|skip_connection|$. In Auto-CsiNet-N6, the opposite is true, demonstrating that NAS can alleviate network degradation to some extent.

Throughout the search process, several patterns in the searched structures were identified. Firstly, the probability of convolutions is higher than $\verb|skip_connection|$, with large-kernel convolutions being more likely to appear than small-kernel ones. This is consistent with the manual design experience, where large kernel sizes enable a large receptive field \cite{guo2022overview,guo2020convolutional, wang2021multi,lu2020multi}. Secondly, the probability of $\verb|zero|$ occurring is almost zero, indicating that NAS tends to assemble operations in parallel rather than in series, as $\verb|zero|$ interrupts information flow in the SuperNet, resulting in a low score. Thirdly, the cell complexity (number of inner nodes $N$) cannot be set to infinity, as overlarge $N$ leads to network degradation, as observed with Auto-CsiNet-N7 performing worse than Auto-CsiNet-N6 due to network degradation.

Figure \ref{fig:architecture characteristic} also reflects the degradation of sub-network performance caused by excessive search. The X-axis indicates the search time point, represented by the number of search epochs, at which the sub-network with optimal performance appears. As the performance of the sub-network fluctuates with the search time, and the appropriate search time varies depending on the scenarios or cell configurations, only a sub-optimal structure can be obtained with a fixed search time. This issue can be addressed through the use of the early stop and elastic selection strategy outlined in Algorithm \ref{alg:Early-stopping}.

\section{Conclusion}\label{section:conclusion}
This paper focuses on the design of NN architectures for intelligent CSI feedback. To address the main challenges in manual design, we propose an automatic generation scheme for NN structures using NAS to substitute the laborious process of adjusting hyperparameters in manual design work. This approach enables a standardized and convenient process for scene-specific customization of NN architectures and integrates implicit scene knowledge from the CSI feature distribution into the architecture design in a data-driven manner.

To reduce the threshold for implementing NAS and its resource consumption in CSI feedback, we employ an efficient gradient descent-based NAS, namely the PC-DARTS method \cite{PC-DARTS}, and control the scale of the search space by constructing it based on CS-CsiNet \cite{wen2018deep}. This helps in controlling and reducing the search cost, and the scheme can be easily implemented in practice. Additionally, we observe that excessive searching leads to a degradation in performance of the search structure. To address this, we further improve the scheme by adopting the early stopping and elastic selection mechanisms.

The resulting searched NN structure, referred to as Auto-CsiNet, outperforms manually designed NN architectures in terms of performance and complexity, validating the effectiveness of the proposed automatic scheme. Furthermore, the scene-specific Auto-CsiNets surpass the manually designed general NN architectures in the given scenes, demonstrating that the proposed scheme achieves scenario-specific customization to maximize performance. The searched NN structure is also consistent with the experience of human design, such as the observation that complex scenes require a larger receptive field. NAS can also alleviate network degradation during the search process if the cell is set to be too complex.

For future work, it is important to consider practical deployment issues. For example, in future communication networks, edge servers will determine the NN architectures for various application scenarios, and limitations such as the power of hardware devices and requirements for latency should be taken into account when selecting suitable candidate operation sets. Depth-separable convolution serves as an example of a candidate operation set that is suitable for lightweight networks and can be easily deployed on mobile devices.

\bibliographystyle{IEEEtran}
\bibliography{reference}

\end{document}